\documentclass[11pt]{article}

\usepackage{acl}
\usepackage{xurl}   
\usepackage{times}
\usepackage{latexsym}
\usepackage[T1]{fontenc}
\usepackage[utf8]{inputenc}
\usepackage{microtype}
\usepackage{inconsolata}
\usepackage{graphicx}
\usepackage{amsmath}
\usepackage{amssymb}
\usepackage{booktabs}
\usepackage{multirow}
\usepackage{xcolor}
\usepackage{soul}

\newcommand{\EDM}{\Delta}
\newcommand{\pvec}{\mathbf{p}}
\newcommand{\fvec}{\mathbf{f}}

\title{Catalyst Papers in Artificial Intelligence Research:\\
A Landscape on ICLR from 2017 to 2025}

\author{
  Fan Huang \\
  Indiana University Bloomington \\
  \texttt{huangfan@acm.org}
}

\begin{document}
\maketitle

\begin{abstract}
A small number of methodological contributions, including word2vec,
the Transformer, large-scale pre-training, and reinforcement learning
from human feedback, have reshaped NLP and AI research over the past
decade. OpenReview now makes numeric reviewer scores and
accept/reject decisions public for every ICLR submission.
Whether such review signals identify trajectory-changing papers at
submission time, however, remains untested at corpus scale. We answer
this question on $36{,}113$ papers from ICLR 2017--2025, identifying
\emph{catalysts}: papers whose descendants measurably redirect future
research. We compare four disruptiveness measures (the
Consolidation/Destabilization (CD) index, node2vec, the
direction-aware Embedding Disruptiveness Measure (EDM), and an
LLM-based semantic rater) and define a five-type operational
catalyst taxonomy (topic
initiator, topic bridge, within-topic redirector, simultaneous, and
recognition-misaligned). EDM leads at identifying highly cited ICLR
papers (AUC $0.83$ vs.\ $0.60$ for CD, $0.49$ for node2vec, and $0.42$
for the LLM rater). Topic initiators precede a $7.55{\times}$
topic-share growth and topic bridges precede an $11.52{\times}$
growth in cross-topic citation flow versus year-matched controls.
We found that the peer review scores are essentially orthogonal to
future disruptiveness ($|\rho|{\leq}0.005$; accepted and rejected
papers have indistinguishable mean EDM, $p{=}0.11$).
\end{abstract}

\section{Introduction}
\label{sec:intro}

A small number of methodological contributions, including word2vec
\citep{mikolov2013word2vec}, the Transformer architecture
\citep{vaswani2017attention}, large-scale pre-training
\citep{radford2018gpt1,radford2019gpt2}, and reinforcement learning
from human feedback
\citep{ouyang2022instructgpt}, have reshaped NLP and AI research over
the past decade. OpenReview makes per-paper reviewer scores and accept/reject
decisions publicly available for every submission to the
International Conference on Learning Representations (ICLR)
\citep{berenslab2025iclr}. Identifying which submissions
later redirect research trajectories has become a central question
in the science of science
\citep{park2023papers,wu2019large,fortunato2018science};
estimating this potential at submission time, however, remains
difficult.

The CD index \citep{funk2017dynamic} captures only one-hop
citation displacement and clusters near zero on sparse networks
\citep{petersen2024disruption,kim2026edm}; whether recent
embedding- or content-based alternatives
\citep{kim2026edm,cohan2020specter}, combined with
peer-review signals, can identify submissions that later reorient a
sub-field remains open.

Prior work has progressed along two largely separate lines. The
bibliometric line introduced the Consolidation/Destabilization (CD)
index \citep{funk2017dynamic},
reported a multi-decade decline in disruptiveness
\citep{park2023papers}, linked team size and atypical combinations
to impact \citep{wu2019large,uzzi2013atypical}, used
citation-structure features to anticipate impact
\citep{clauset2017data}, and recently proposed the Embedding
Disruptiveness Measure (EDM), a direction-aware alternative that is
more robust to sparse networks \citep{kim2026edm}. The peer-review
line has documented status effects
\citep{merton1968matthew,teplitskiy2022status}, novelty penalties
\citep{wang2017value}, and reviewer inconsistency at ML venues
\citep{cortes2021inconsistency}.

The two lines have rarely been joined: journal corpora (Web of
Science, APS) do not include per-paper reviewer scores
\citep{park2023papers,kim2026edm,clauset2017data}, while
conference-side work has not connected reviewer signals to long-run
trajectory change \citep{cortes2021inconsistency}. Prior work has examined the relationship between ML-conference
review scores and raw citation outcomes
\citep{tran2020openreview}, but has not extended this to
direction-aware, multi-generational trajectory measures. Whether submission-time review signals predict long-run
trajectory change at a major ML venue therefore remains open.

We define a \emph{catalyst paper} as a submission whose descendants
measurably redirect later research. We address this gap with three
connected research questions on the nine-year ICLR record
(2017--2025). RQ1 (Measurement) asks which
operationalization best identifies catalyst papers: topology-based
metrics \citep{funk2017dynamic}, generic graph embeddings
\citep{grover2016node2vec}, the directional citation embedding EDM
\citep{kim2026edm}, or an LLM-based semantic rater. RQ2 (Mechanisms)
asks through what mechanisms
catalyst papers reshape ICLR research, considering topic initiation,
topic bridging, within-topic redirection, and simultaneous
discovery. RQ3 (Recognition) asks whether submission-time reviewer
signals track future trajectory change, and where the residual
miscalibration concentrates.

This paper makes four contributions: (i) a formal five-type catalyst
taxonomy; (ii) a head-to-head comparison of four disruptiveness
measures on a common ICLR corpus; (iii) year-matched evidence that catalyst
types precede subsequent topic-share growth and cross-topic
citation-flow changes; and (iv) the first analysis pairing per-paper
OpenReview signals with direction-aware, multi-generational
trajectory measures (EDM), finding $|\rho| \leq 0.005$ between
reviewer scores and EDM and a percentile gap structured by catalyst
type and by topic.

\section{Related Work}
\label{sec:related}

\paragraph{Disruption measures.}
The Consolidation/Destabilization (CD) index
\citep{funk2017dynamic} flags a paper as disruptive when later citers
ignore its references and has anchored arguments about a multi-decade
decline in disruptiveness \citep{park2023papers}.
Critiques show CD to be bimodally degenerate on sparse networks,
blind to multi-generational influence \citep{kim2026edm}, and biased
by citation inflation \citep{petersen2024disruption}. The Embedding
Disruptiveness Measure (EDM) of \citet{kim2026edm} addresses these
limitations via direction-aware random walks and outperforms CD on
milestone-paper identification in Web of Science and APS.
Content-side alternatives such as SPECTER and SciRepEval
\citep{cohan2020specter,singh2022specter2} motivate the LLM-rater
family.

\paragraph{Science of science in NLP and ML.}
Most empirical work in the science of science
\citep{fortunato2018science,uzzi2013atypical,clauset2017data}
derives from journal corpora with multi-decade timescales and private
peer review, often analyzed through Web of Science or OpenAlex
\citep{priem2022openalex}. The NLP and ML conference setting differs:
arXiv-mediated diffusion compresses idea propagation to months,
OpenReview makes per-paper reviewer scores public for every
submission, and rejected papers often persist via preprint servers
\citep{berenslab2025iclr}.

\paragraph{Peer review and recognition.}
Status biases \citep{merton1968matthew,teplitskiy2022status} and
novelty penalties \citep{wang2017value} are well documented in
journal peer review. At ML and NLP venues, the NeurIPS consistency
experiment found a $25.9\%$ rate of inconsistent accept/reject
decisions across two independent committees reviewing the same $166$
submissions \citep{cortes2021inconsistency}; an ICLR-specific
analysis of the 2017--2020 window paired OpenReview scores with
Semantic Scholar citation impact and reported only weak
score--citation correlation across both accepted and rejected
submissions \citep{tran2020openreview}.

\section{Data}
\label{sec:data}

Our corpus is the ICLR submission record 2017--2025 from the
\texttt{berenslab/iclr-dataset} \citep{berenslab2025iclr}, containing
$36{,}113$ papers across nine years with OpenReview identifiers, titles,
abstracts, authors, acceptance decisions, and integer reviewer scores
(1--10). Submissions grow from $489$ (2017) to $11{,}663$ (2025);
accept rates are stable at $26$--$33\%$. Per-year counts are reported
in Table~\ref{tab:data} (Appendix~\ref{app:network}). For each paper
we derive per-paper review summaries (mean $\bar{s}$, variance
$\sigma^2_s$, number of reviewers $n_r$); papers with no recorded
scores are excluded from peer-review analyses.

\paragraph{Citation network.}
We construct a directed citation graph $G=(V,E)$ over the 36{,}113 ICLR
papers. We match each paper to a Semantic Scholar (S2) record via title-based
fuzzy matching (RapidFuzz token-set ratio $\geq 0.85$, year tolerance
$\pm 2$) and retrieve reference lists through the S2 Graph API, keeping
only ICLR-internal edges. We use S2 rather than OpenAlex because S2
indexes arXiv preprints directly, giving more complete coverage of ML
and NLP submissions. We match $27{,}843$ papers ($77.1\%$); unmatched
submissions are concentrated among rejected papers that were never
reposted to arXiv.

\section{Methods}
\label{sec:methods}

\subsection{Catalyst Taxonomy}
\label{sec:taxonomy}

We define a \emph{catalyst paper} as one whose descendants measurably
reshape the topic or citation structure of subsequent research.
Catalyst status sits within the broader empirical tradition of
identifying high-impact papers in citation networks
\citep{wu2019large,uzzi2013atypical,clauset2017data}. As a further
operational investigation of the \emph{disruptive paper} concept
\citep{kim2026edm,funk2017dynamic}, we operationalize catalyst
status as a multi-label family of five threshold-defined mechanism
types, computed over the ICLR-internal citation graph and topic
assignments from Section~\ref{sec:topics-validation}.

\paragraph{Topic Initiator (TI).}
A new topic cluster appears, or an existing one grows at an
accelerated rate, in the following 1--3 years; this parallels
sub-field-birth events studied in
\citet{kim2026edm,uzzi2013atypical}. Let $\text{share}(T, y)$ be
the fraction of ICLR papers in year $y$ assigned to topic $T$, and
define
\begin{align}
  \bar{s}^{\text{pre}}_i  &= \tfrac{1}{2}\!\sum_{y=y_i-2}^{y_i-1} \text{share}(T_i, y), \\
  \bar{s}^{\text{post}}_i &= \tfrac{1}{3}\!\sum_{y=y_i+1}^{y_i+3} \text{share}(T_i, y).
\end{align}
Paper $i$ is a TI if $\bar{s}^{\text{post}}_i / \bar{s}^{\text{pre}}_i \geq 2.0$.

\paragraph{Topic Bridge (TB).}
The paper is followed by a measurable increase in cross-topic
citation flow between previously weakly connected clusters,
analogous to brokerage across structural holes
\citep{burt1992structural} and to the atypical-combination
mechanism behind high-impact work \citep{uzzi2013atypical}. Let
$F_i$ be the cross-topic citation flow attributed to paper $i$'s
topic cluster in its publication year, and $D_i$ the number of
distinct topic clusters reached by $i$'s citation descendants
within two years. Paper $i$ is a TB if $F_i$ falls in the top
$10\%$ of the cross-topic flow distribution and $D_i \geq 2$.

\paragraph{Within-Topic Redirector (WR).}
The topic label is preserved, but the embedding centroid of the
topic's subsequent papers shifts substantially. WR is the
within-field analog of the CD-index notion of disruption
\citep{funk2017dynamic}, but uses direction-aware embeddings to
avoid the sparse-network and citation-inflation biases of CD
\citep{petersen2024disruption,kim2026edm}. Let $\mathbf{c}(T, y)$
be the mean embedding of papers in topic $T$ at year $y$; the
centroid shift induced by paper $i$ is
$\|\mathbf{c}(T_i, y_i) - \mathbf{c}(T_i, y_i-1)\|$, $z$-scored
within $T_i$. Paper $i$ is a WR if this $z$-score exceeds the
cluster-conditional median.

\paragraph{Simultaneous Catalyst (SC).}
The future-vector neighborhood clusters with contemporaneous papers,
with no single paper claiming the breakthrough. SC operationalizes
the multiple-discovery phenomenon documented since
\citet{merton1961singletons} and detected via citation embeddings
by \citet{kim2026edm}; the full identification procedure and
empirical results are in Section~\ref{sec:rq2-simul}.

\paragraph{Recognition-Misaligned (RM).}
High retrospective $\EDM$ paired with weak, borderline, or contested
submission-time review signals. RM captures a submission-stage
analog of the \emph{sleeping-beauty} phenomenon
\citep{ke2015sleeping}, consistent with novelty penalties in peer
review \citep{wang2017value} and the only weak score-to-citation
correlation reported at ICLR \citep{tran2020openreview}. Paper $i$
is an RM if (i) $\EDM_i$ is at or above the 90th percentile of the
full-corpus EDM distribution ($\EDM \geq 0.899$); and (ii) its
reviewer mean score $\bar{s}_i$ is at or below the year-conditional
acceptance boundary minus $0.5$ points, or its reviewer score
variance $\sigma^2_{s,i}$ places it in the ``controversial''
category. In practice, $83.7\%$ of RM papers were rejected at ICLR.

\paragraph{Why these five types.}
A single disruptiveness scalar collapses qualitatively distinct
mechanisms into one number; recent work has shown such scalars to
be bimodally degenerate, citation-inflation biased, and blind to
multi-generational structure
\citep{petersen2024disruption,kim2026edm}. The science-of-science
literature has correspondingly identified multiple distinct routes
to high impact rather than a single one, including small-team
disruption versus large-team development \citep{wu2019large} and
the combination of conventional with atypical references
\citep{uzzi2013atypical}. Our five-type taxonomy operationalizes
this mechanism plurality: sub-field birth (TI), brokerage (TB),
within-field reshaping (WR), simultaneous breakthroughs (SC), and
review-signal misalignment (RM) each map to a separate literature
above and apply as multi-label rather than mutually exclusive
categories. The threshold-based definitions are also
venue-portable, supporting comparisons with other ML and NLP
corpora where existing review-impact analyses have so far been
restricted to a single venue at a time
\citep{tran2020openreview,cortes2021inconsistency}.

\subsection{Measuring Disruptiveness}
\label{sec:measures}

We compare four disruptiveness measures on the hypothesis that
topology, graph embeddings, and LLMs access distinct information
channels.

\paragraph{(M1) CD index.}
The disruption index $D_i = (n_f - n_b)/(n_f + n_b + n_k)$ of
\citet{funk2017dynamic}, where $n_f, n_b, n_k$ are the standard citation
neighborhood counts.

\paragraph{(M2) Generic graph embedding: node2vec.}
Undirected node2vec \citep{grover2016node2vec} on $G$ with the same
hyperparameters as EDM ($T{=}160$, $R{=}80$, $d{=}100$, $c{=}5$). The score
$\text{M2}_i$ is cosine distance between the mean reference and mean citer
embeddings of paper $i$.

\paragraph{(M3) Directional citation embedding: EDM.}
Following \citet{kim2026edm}, we learn a past vector $\pvec_i$ and future
vector $\fvec_i$ for each paper $i$ from direction-aware random walks
trained with single-side skip-gram context. The score is
\begin{equation}
  \EDM_i = 1 - \frac{\fvec_i \cdot \pvec_i}{|\fvec_i|\,|\pvec_i|}
  \label{eq:edm}
\end{equation}
We use the canonical hyperparameters $T{=}160$, $R{=}80$, $d{=}100$,
$c{=}5$, $\kappa^{\text{in}}_u$ in-degree weighting. The training
objective, full equations, and a 1-D parameter sweep showing ERS AUC
is stable in $[0.72,0.76]$ across $T \in \{80,160,240\}$ and
$d \in \{64,100,128\}$ are in Appendices~\ref{app:methods-extended} and
\ref{app:params}.

\paragraph{(M4) LLM rater.}
For each paper, we prompt \texttt{gpt-4o-mini}
\citep{openai2024gpt4omini} with the paper's title and abstract and
request a single integer disruption-potential score on a $0$--$10$
scale, with no other text. Scoring is zero-shot (no fine-tuning) and
deterministic ($\tau{=}0$); M4 covers the full $36{,}113$-paper
corpus. The rubric distinguishes disruption from paper quality,
novelty, and citation count (full prompt in
Appendix~\ref{app:llm-prompt}). M4 sees only paper content (title
and abstract); it never reads the citation graph or reviewer scores.

\subsection{Topic Modeling and Validation Sets}
\label{sec:topics-validation}

\paragraph{Topics.} We embed each paper's \texttt{title [SEP] abstract} via
OpenAI \texttt{text-embedding-3-large} \citep{openai2024textembed} at
$1{,}024$ dimensions, project to 50-D with UMAP \citep{mcinnes2018umap},
and cluster with HDBSCAN \citep{campello2013density}
($\text{min\_cluster\_size}{=}50$). This yields 113 topic clusters
(12{,}102 noise papers; $33.5\%$). Hyperparameters and the topic-share
$\text{share}(T,y)$ formula are in Appendix~\ref{app:methods-extended}.

\paragraph{Validation sets.} We evaluate each measure against four
signals. (1) External Recognition Set (ERS), the top $2\%$ by
ICLR-internal citation count ($n{=}739$ positives), capturing structural
recognition by the field. (2) LLM-judge Annotation Set (LAS), a stratified
random sample of $50$ papers labeled by two independent runs of an
LLM judge (\texttt{claude-opus-4-6}) over title and abstract (union of
positive labels: $9$ positives, $41$ negatives; run-to-run agreement
$\kappa{=}0.291$; Appendix~\ref{app:annotation}). LAS measures
cross-model semantic-rubric agreement, not agreement with human
expert judgment.
(3) Citation velocity, the Spearman correlation between each measure
and citations per year since publication. (4) Twenty-seven ICLR Best
Paper winners (2021--2025) used as qualitative case studies. We report
ROC-AUC and odds ratios from Firth's penalized logistic regression
\citep{firth1993bias} per $10$-percentile increase. The citation-count
baseline is excluded from ERS because it is definitionally circular.
Details and stratification design are in Appendix~\ref{app:validation}.

\section{RQ1: Measuring Catalyst Papers}
\label{sec:rq1}

\begin{figure*}[t]
  \centering
  \includegraphics[width=\linewidth]{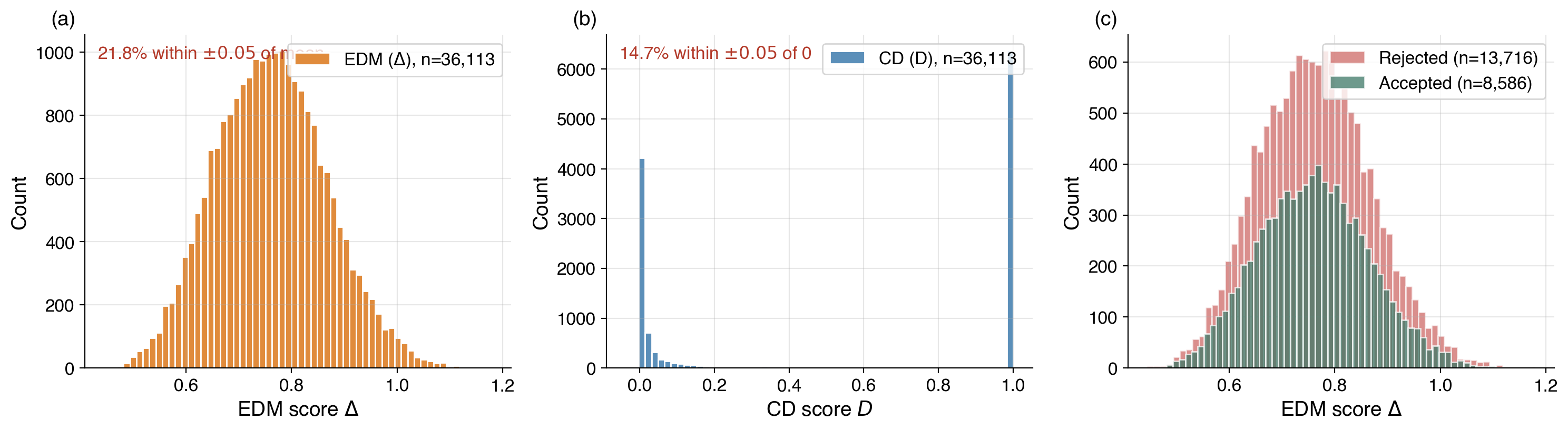}
  \caption{Disruptiveness distributions across $36{,}113$ ICLR papers.
  (a) EDM is smooth and near-Gaussian
  ($\bar{\Delta}{=}0.76$, std $0.11$, range $[0.44,1.18]$).
  (b) CD is bimodally degenerate, with scores concentrated near $0$ and
  near $1$.
  (c) EDM distributions are nearly identical for accepted and rejected
  papers.}
  \label{fig:dist}
\end{figure*}

To compare measure coverage and distribution shape on the ICLR corpus,
we apply CD, node2vec, EDM, and the LLM rater to all $36{,}113$ papers
and inspect the per-measure distributions
(Fig.~\ref{fig:dist}). EDM produces a smooth, near-Gaussian
distribution and covers $22{,}302$ papers ($62\%$); CD covers $35\%$
and is bimodally degenerate (Fig.~\ref{fig:dist}b); node2vec covers
$18\%$; the LLM rater (M4) covers the full corpus. Among the eight
ICLR 2024--2025 Best Paper winners, only one receives a CD score
while all receive EDM and LLM scores, and the temporal trend of
$\EDM$ is essentially flat across 2017--2025 (year-wise mean ranges
from $0.747$ in 2024 to $0.823$ in 2017; year-wise std stays within
$[0.101, 0.125]$). The CD bimodality reflects the sparsity of the
ICLR-internal citation network relative to the Web of Science and APS
networks used in prior work
\citep{funk2017dynamic,park2023papers,kim2026edm}, and the flat EDM
trend does not reproduce the monotonic decline reported by
\citet{park2023papers} for broad science. Extended analysis and the
yearly trend figure are in Appendix~\ref{app:rq1-extended}.

\subsection{Validation Against Benchmark Sets}
\label{sec:rq1-valid}

\begin{table}[ht]
  \centering
  \small
  \begin{tabular}{lcc}
    \toprule
    Measure & LAS AUC & ERS AUC \\
    \midrule
    M1: CD Index          & 0.566 & 0.596 \\
    M2: node2vec          & 0.276 & 0.493 \\
    M3: EDM               & 0.496 & 0.827 \\
    M4: LLM               & 0.749 & 0.424 \\
    Citations (baseline)  & 0.526 & /\footnotemark{} \\
    \bottomrule
  \end{tabular}
  \caption{ROC-AUC for each measure against the LLM-judge annotation
  set (LAS, union of two independent \texttt{claude-opus-4-6} runs,
  $9$ positives) and external citation recognition (ERS, top $2\%$
  by citation count, $739$ positives). EDM leads on structural
  validation, while the LLM rater leads on cross-model semantic-rubric
  agreement.}
  \label{tab:auc}
\end{table}
\footnotetext{Circular: ERS is defined as the top 2\% by citation count, so this cell
would trivially be 1.000.}

\begin{table}[ht]
  \centering
  \small
  \resizebox{\linewidth}{!}{%
  \begin{tabular}{llccc}
    \toprule
    Measure & Set & OR & 95\% CI & $p$ \\
    \midrule
    M3: EDM           & ERS & 1.744 & [1.671, 1.821] & $<0.001$ \\
    M1: CD Index      & ERS & 1.156 & [1.121, 1.192] & $<0.001$ \\
    M4: LLM           & ERS & 0.901 & [0.876, 0.926] & $<0.001$ \\
    M2: node2vec      & ERS & 0.991 & [0.956, 1.028] & $0.63$ \\
    \midrule
    M4: LLM           & LAS & 1.414 & [1.041, 1.920] & $0.027$ \\
    M1: CD Index      & LAS & 1.078 & [0.837, 1.390] & $0.56$ \\
    Citations         & LAS & 1.029 & [0.806, 1.314] & $0.82$ \\
    M3: EDM           & LAS & 0.996 & [0.761, 1.304] & $0.98$ \\
    M2: node2vec      & LAS & 0.779 & [0.530, 1.145] & $0.20$ \\
    \bottomrule
  \end{tabular}}
  \caption{Firth logistic regression odds ratios (OR) per
  $10$-percentile increase in each measure, for ERS and LAS
  membership. OR~$>1$ indicates that a 10-percentile increase in the
  measure multiplies the odds of set membership; OR~$<1$ indicates a
  negative association. The EDM effect size on ERS, measured as the
  odds-ratio deviation from $1$, is $4.77\times$ that of CD
  ($(1.744-1)/(1.156-1) = 4.77$), and the LLM rater is the only
  measure significantly associated with LAS at $\alpha = 0.05$.}
  \label{tab:firth}
\end{table}

We evaluate each disruptiveness measure on two validation signals,
the External Recognition Set (ERS, top $2\%$ by ICLR-internal
citation count) and the LLM-judge Annotation Set (LAS), reporting
ROC-AUC and Firth-regression odds ratios
(Tables~\ref{tab:auc},~\ref{tab:firth}). EDM achieves ERS AUC
$0.827$ versus $0.596$ for CD and $0.493$ for node2vec, with each
$10$-percentile increase in EDM multiplying the odds of top-cited
membership by $1.74$ (versus $1.16$ for CD), while node2vec is
statistically indistinguishable from random ($p{=}0.63$). M4 (LLM)
shows the inverse pattern, leading on LAS (AUC $0.749$, OR $1.41$,
$p{=}0.03$) but performing below chance on ERS ($0.424$). The
contrast between M2 and M3 isolates direction-aware walk training as
the salient methodological ingredient on the structural side; since
LAS labels come from an independent LLM judge
(\texttt{claude-opus-4-6}), M4's LAS lead measures cross-model
agreement on the same semantic rubric, not validation against expert
judgment. EDM and the LLM rater are
near-orthogonal on both validation signals.

\paragraph{NLP-subset replication.} Restricting the analysis to
NLP- and language-modeling topic clusters ($11{,}255$ papers across
$55$ clusters covering language modeling, embeddings, and
vision-language work) yields an EDM ERS AUC of $0.860$, compared with
$0.827$ on the full corpus; CD AUC drops to $0.597$ and node2vec to
$0.499$. The RQ1 and RQ3 conclusions therefore appear to hold on the
NLP-focused subset (Appendix~\ref{app:rq1-extended}).

To probe complementarity beyond AUC, we compute citation velocity
(the Spearman correlation between each measure and citations per
year since publication) and examine the ICLR 2021--2025 Best Paper
winners as case studies. Citation velocity is consistent with the
ERS ranking: EDM $\rho{=}{+}0.300$, CD $\rho{=}{+}0.127$, node2vec
$\rho{=}{-}0.293$, LLM $\rho{=}{+}0.016$ (ns). The case studies
expose two complementary blind spots: EDM tends to under-rank
canonical follow-ups (for example, Score-Based SDE at the $26.4$th
percentile and Analytic-DPM at the $8.4$th), whose descendants
remain close to their antecedents, while the LLM rater saturates at
the top of its scale, assigning $7/10$ (the $83.8$th percentile) to
most Best Paper winners and $6/10$ or below to the rest. Structural
and semantic measures therefore appear to tap different signals,
each with its own failure mode. Detailed case studies, correlation
heatmaps, and scatter plots of divergent papers are in
Appendix~\ref{app:rq1-extended}.

\begin{table}[t]
\centering\footnotesize
\setlength{\tabcolsep}{4pt}
\begin{tabular}{@{}lrr@{}}
\toprule
Model pair & Spearman $\rho$ & Quad.\ $\kappa$ \\
\midrule
gpt-4o-mini vs.\ gpt-4o            & 0.71 & 0.48 \\
gpt-4o-mini vs.\ claude-sonnet-4-6 & 0.56 & 0.23 \\
gpt-4o-mini vs.\ llama-3.3         & 0.68 & 0.29 \\
gpt-4o-mini vs.\ qwen-2.5          & 0.55 & 0.34 \\
gpt-4o vs.\ claude-sonnet-4-6      & 0.63 & 0.52 \\
gpt-4o vs.\ llama-3.3              & 0.72 & 0.16 \\
gpt-4o vs.\ qwen-2.5               & 0.57 & 0.16 \\
claude-sonnet-4-6 vs.\ llama-3.3   & 0.56 & 0.07 \\
claude-sonnet-4-6 vs.\ qwen-2.5    & 0.46 & 0.07 \\
llama-3.3 vs.\ qwen-2.5            & 0.54 & 0.45 \\
\midrule
Mean                               & 0.60 & 0.28 \\
\bottomrule
\end{tabular}
\caption{Inter-LLM pairwise agreement on a stratified
$2{,}000$-paper sample, across five models from four vendors (OpenAI,
Anthropic, Meta, Alibaba). All $\rho$ are highly significant
($p<10^{-100}$).}
\label{tab:llm-agreement}
\end{table}

EDM is the strongest single signal (AUC $0.83$ vs.\ $0.60/0.49/0.42$
for CD/node2vec/LLM); its blind spots complement the LLM rater's.

\section{RQ2: Mechanisms of Trajectory Change}
\label{sec:rq2}

\paragraph{Topics and catalyst counts.}
HDBSCAN over the $50$-dimensional UMAP of text embeddings yields $113$
topic clusters covering $66.5\%$ of papers (the largest $10$ clusters
table and full list are in Appendix~\ref{app:rq2-extended}). The four
operational catalyst types are not mutually exclusive: TI~$=3{,}063$
($8.5\%$), TB~$=3{,}539$ ($9.8\%$), WR~$=2{,}562$ ($7.1\%$), and
RM~$=1{,}119$ ($3.1\%$), with a union of $8{,}015$ papers ($22.2\%$).
The co-occurrence structure itself is informative: $26.4\%$ of TI
papers are also WR, $36.6\%$ of RM papers are also TB (reviewer
calibration appears weakest on bridges), but only $3.7\%$ of TI papers
are RM. Operational thresholds and the full $4{\times}4$ co-occurrence
matrix are in Appendix~\ref{app:rq2-extended}.

\subsection{Topic Dynamics Driven by Catalyst Papers}
\label{sec:rq2-topic}

To estimate the topic-dynamic effect of TI papers we compare each TI
paper's topic-share trajectory to a year-matched control of up to 5
non-TI papers in different topic clusters (holding cohort effects
constant); for TB papers we compute the change in cross-topic
citation flow into their cluster over the $\pm 2 / +3$-year window.
TI papers precede topic-share growth at $7.55\times$ the rate of
matched controls (Table~\ref{tab:ti-growth}; Welch $t{=}46.06$,
$p{<}10^{-300}$); for TB papers, the mean cross-topic citation flow
into their cluster grows from $242.3$ to $848.5$ edges per year, an
$11.52\times$ mean multiplicative increase (median $4.36\times$;
IQR $[2.58, 9.34]$; $n{=}3{,}228$). Propensity matching on
log-citation count, acceptance, and year (1-NN, caliper $0.05$;
$n{=}3{,}063$ matched) reduces the TI ratio to $6.31\times$ (Welch
$t = 44.0$, $p < 10^{-300}$; Appendix~\ref{app:rq2-extended}); the
$16\%$ drop from $7.55\times$ to $6.31\times$ quantifies the
citation-count confound, leaving a large and significant residual.
In effect-size terms TB is the largest catalyst mechanism observed
in the corpus, suggesting that bridging papers warrant at least as
much attention as topic-initiating ones. The full TB flow table,
distribution figure for TI versus controls, and cross-topic flow
heatmap are in Appendix~\ref{app:rq2-extended}.

\subsection{Simultaneous Discoveries and Cross-Domain Diffusion}
\label{sec:rq2-simul}

To identify simultaneous catalysts, we follow \citet{kim2026edm} and
select same-year, no-author-overlap pairs with future-vector cosine
$\geq 0.9$, then filter for descendant support ($\geq 3$
ICLR-internal citations) and valid topic clusters; a citation-graph
ground-truth check uses co-citation overlap $\geq 0.20$ together
with the absence of a direct edge. The $483{,}809$-pair raw pool
collapses to $162$ candidates after filtering, of which only $4$
($2.5\%$) are same-topic pairs; the ground-truth check confirms
$76.25\%$ precision at top-80 but reports only $0.20\%$ prevalence
in the raw pool. Simultaneous discoveries therefore appear
substantially rarer in AI than in the physics setting of
\citet{kim2026edm}, consistent with arXiv preprint culture
compressing parallel discoveries into sequential citation chains.
Filtering funnel, threshold sensitivity, and the one confirmed
same-topic pair are in Appendix~\ref{app:rq2-extended}.

For cross-domain diffusion analysis, we retrieve $792{,}018$
in-window citing works for $7{,}332$ ICLR papers via the S2 Graph
API and compute two metrics on the citing-paper side: a
\emph{composition} metric (the fraction of a paper's citers that
are non-AI) and a \emph{reach} metric (the fraction of papers in a
catalyst class with any non-AI citer). Composition places TB and WR
above the non-catalyst baseline (Mann--Whitney $p \leq 0.004$),
whereas reach places TI at the top of every non-AI domain (Biology
$17\%$, Healthcare $52\%$, other CS $58\%$); RM papers are lowest on
both metrics. The two metrics therefore tell different stories: TB
and WR papers attract a higher \emph{share} of citations from
outside AI, while TI papers more uniformly succeed in attracting
\emph{any} non-AI citers across diverse domains, and RM's low
performance on both is consistent with the under-recognition
pattern reported in RQ3. The full methodology, figures, and limitations
of the S2 \texttt{fieldsOfStudy} taxonomy are in
Appendix~\ref{app:rq2-extended}.

Topic bridges drive the largest catalyst effect ($11.52\times$
cross-topic flow); topic initiators precede $7.55\times$ sub-field
growth; simultaneous discoveries are an order of magnitude rarer in
AI than in physics.

\section{RQ3: Peer Review and Recognition}
\label{sec:rq3}

\subsection{Review Scores and Future Disruptiveness}
\label{sec:rq3-scores}

To test whether submission-time review signals predict future
disruptiveness, we compute Spearman correlations between four review
features (mean score $\bar{s}$, variance $\sigma^2_s$, range, and
number of reviewers $n_r$) and EDM, fit Firth logistic regressions
predicting top-decile $\EDM$ membership
(Table~\ref{tab:firth-review}), and compare accepted versus rejected
papers in the citation network via Mann--Whitney. The Spearman
correlations lie within $|\rho| \leq 0.005$; only the reviewer count
reaches nominal significance ($\rho = -0.045$, $p{<}0.001$), and it
explains less than $0.2\%$ of the variance. Firth odds ratios stay
within $2\%$ of $1.0$, none significant at $\alpha{=}0.05$. Accepted
and rejected papers ($n_{\text{acc}}{=}8{,}586$;
$n_{\text{rej}}{=}13{,}716$) have indistinguishable mean EDM
($0.7597$ vs.\ $0.7624$; Mann--Whitney $p{=}0.11$), and rejected
papers are slightly over-represented in the top-decile of EDM
($10.3\%$ vs.\ $9.5\%$). At the corpus level, ICLR peer review
therefore appears largely orthogonal to future disruptiveness as
captured by EDM. The violin plot of EDM by score quartile and the
reviewer-bin heatmap are in Appendix~\ref{app:rq3-extended}.

\begin{table}[ht]
\centering
\footnotesize
\resizebox{\linewidth}{!}{%
\begin{tabular}{lrrrrr}
\toprule
Feature & $\rho$ & $p$ & OR & 95\% CI & $p_{\text{F}}$ \\
\midrule
Mean score ($\bar{s}$)        & $-0.003$ & $0.62$   & $0.995$ & $[0.979, 1.010]$ & $0.52$ \\
Score var.\ ($\sigma^2_s$)    & $-0.005$ & $0.49$   & $0.999$ & $[0.983, 1.015]$ & $0.90$ \\
Score range                   & $-0.003$ & $0.69$   & $1.000$ & $[0.984, 1.016]$ & $0.97$ \\
\# reviewers ($n_r$)          & $-0.045$ & $<0.001$ & $0.980$ & $[0.959, 1.002]$ & $0.08$ \\
\bottomrule
\end{tabular}}
\caption{Spearman $\rho$ and Firth odds ratio (OR; per 10-pctile
increase) predicting top-$\EDM$-decile; $p_{\text{F}}$ = Firth $p$.
All OR within 2\% of 1.0; none significant at $\alpha{=}0.05$.
$n=8{,}586$ accepted papers.}
\label{tab:firth-review}
\end{table}

\subsection{Structured Miscalibration}
\label{sec:rq3-miscal}

We define the \emph{review gap} for a paper as its EDM percentile
minus its reviewer-score percentile, with a positive gap indicating
that future disruption exceeded what the review score would have
predicted, and compute the mean gap by catalyst type with $t$-tests
against non-catalysts (Table~\ref{tab:miscal-type}). RM papers show
a mean gap of $+0.59$ ($t = 85.75$, $p<0.001$), spanning more than
half of the percentile range; TB papers show a mean gap of $+0.15$
($t = 33.97$, $p<0.001$); TI papers are near zero ($-0.095$;
$p=0.37$, ns); and WR papers show a small negative gap ($-0.003$;
$p<0.001$). Reviewers therefore appear well calibrated on
within-topic work but systematically under-score cross-topic and
recognition-misaligned contributions, which are the catalyst types
most associated with redirecting research trajectories in this
corpus. The full per-type table is in
Appendix~\ref{app:rq3-extended}.

To assess what happens to ICLR-rejected papers, we track their
reappearance on arXiv and other indexed venues; for those that do
reappear we compare EDM distributions against accepted papers,
separately examine borderline rejections (within $0.5$ score points
of the year's median-accepted score) versus clear rejects via
external citation counts, and look at the subset rejected at one
cycle and re-accepted at a subsequent ICLR cycle. Of $24{,}900$
rejected ICLR submissions, $71.2\%$ never reappear on arXiv or any
other indexed venue. The $7{,}167$ that do reappear have mean
$\EDM$ statistically indistinguishable from accepted papers
(Mann--Whitney $p{=}0.11$) and are slightly over-represented in the
top-decile of $\EDM$ ($10.3\%$ vs.\ $9.5\%$), the top $5\%$
($5.2\%$ vs.\ $4.8\%$), and the top $1\%$ ($1.1\%$ vs.\ $0.8\%$);
borderline rejects accumulate substantially more external citations
than clear rejects (median $\log(1{+}\text{cit})$ $2.30$ vs.\
$1.79$, $p{<}0.001$; robust to
$\delta \in \{0.25,0.50,0.75,1.00\}$), and the $280$ papers
rejected at one ICLR cycle and later accepted at a subsequent ICLR
cycle have median $\EDM$ essentially identical to never-rejected
papers ($0.758$ vs.\ $0.759$, $p{=}0.61$). The picture is therefore
bimodal: most rejected submissions disappear from the scholarly
record, but those that survive look statistically similar to
accepted papers on EDM-based trajectory measures, with the
borderline-reject subset only modestly under-credited in external
citation counts. Full sensitivity tables and a four-panel
rejected-papers figure are in Appendix~\ref{app:rq3-extended}.

\subsection{Topic-Dependent Biases}
\label{sec:rq3-topic}

To test for topic-level miscalibration, we compute the review gap
within each topic cluster and rank clusters by their mean gap
(Table~\ref{tab:topic-bias}, Appendix~\ref{app:rq3-extended}). The
most over-valued clusters are Text-to-Video Generation (gap
$-0.265$), Masked Image Modeling ($-0.260$), Vision Transformers
($-0.249$), State Space Sequence Modeling ($-0.225$), and Diffusion
($-0.220$); the most under-valued are Quantum ML ($+0.193$), Active
Learning ($+0.093$), Safe RL ($+0.090$), and Federated Learning
($+0.050$). Reviewers in this corpus therefore tend to reward
fashionable sub-fields beyond the future-disruption signal captured
by EDM and to penalize niche or methodologically unusual areas;
controlling for catalyst type does not eliminate the topic-gap
signal, suggesting that topic bias and catalyst-type bias are
largely independent dimensions of miscalibration. The visual
summary of topic gaps is in Appendix~\ref{app:rq3-extended}.

Review signals are essentially orthogonal to long-run trajectory
change at ICLR ($|\rho|\leq 0.005$); the residual miscalibration is
structured by catalyst type and topic, not random reviewer noise.

\section{Discussion and Limitations}
\label{sec:discussion}

Three properties of the ICLR record enable the present corpus-level
comparison: public per-paper OpenReview scores, a preprint ecosystem
in which $28.8\%$ ($7{,}167$ of $24{,}900$) of rejected submissions
remain visible, and a nine-year window for multi-generational
citation chains. The comparison is essentially null: reviewer signals
do not track future EDM, and the residual miscalibration is
structured by catalyst type and by topic rather than by reviewer
noise. We do not interpret this as evidence that peer review has
failed; a more cautious reading is that conference gatekeeping
selects on rigor, immediate contribution, and fit, properties
partially orthogonal to long-run trajectory change.

Several limitations bound the interpretation of these findings. The
citation graph is restricted to ICLR-internal edges, so submissions
rejected from ICLR and never re-indexed elsewhere are not observed;
sparse citation coverage for the 2024--2025 cohorts narrows the EDM
window for the most recent two years. Per-paper EDM ranks are sensitive
to the walk-generation pipeline, though distribution-level conclusions
are stable across hyperparameters and seeds
(Appendix~\ref{app:params}). The LLM-judge Annotation Set (LAS) is
modest in size ($n{=}50$, between-run $\kappa = 0.291$ for the LLM
judge), which widens the confidence intervals on the LLM rater's LAS
odds ratio; inter-rater agreement across five models from four
vendors is reported on a stratified $2{,}000$-paper subsample
(Appendix~\ref{app:llm-prompt}). Further limitations on the EDM
canonical-follow-up blind spot, simultaneous-discovery validation, and
citation-network coverage are detailed in
Appendix~\ref{app:limitations-extended}.

\section{Conclusion and Future Work}
\label{sec:conclusion}

This work introduces an operational catalyst taxonomy and compares
four disruptiveness measures across nine years of ICLR submissions,
with corresponding reviewer signals linked to each paper. The
directional citation embedding (EDM) shows the highest agreement
with external structural recognition (ERS AUC $0.83$), while the LLM
rater best matches an independent LLM-judge baseline (LAS AUC
$0.75$); topic-initiator catalysts precede a $7.55\times$ growth in
topic share relative to year-matched controls, and topic-bridge
catalysts precede an $11.52\times$ growth in cross-topic citation
flow.

These results suggest that catalyst signals at ICLR are at least
partly separable from review-time signals, and that
program-committee design may benefit from explicit consideration of
trajectory-changing potential alongside the immediate contribution
properties that reviewers currently optimize for.

Several directions extend this work. A multi-venue replication
(NeurIPS, ACL, EMNLP, ICML) would test how much of the recognition
gap is conference-specific. Scoring with additional current LLM
snapshots would broaden the LLM-rater claim. Collecting human expert
annotations on the disruption-label task would let LAS serve as a
human-validated gold standard rather than a cross-model
semantic-rubric baseline. Finally, a longitudinal re-evaluation that
revisits each paper's EDM trajectory would let us study the dynamics
of catalyst recognition rather than only a static snapshot.

\section*{Ethics Statement}

This work analyzes a publicly available scholarly dataset
(\texttt{berenslab/iclr-dataset}) of ICLR submissions and OpenReview
records; no private review identities are used, and all scores and
decisions are already public on OpenReview at the time of analysis.
The dataset comprises publicly-posted submission metadata (titles,
abstracts, author lists, reviewer scores, accept/reject decisions).
Author names and affiliations appear in the source records but are
not surfaced in our analyses, which report results in aggregate or
by paper ID, never by author. Reviewer scores are numeric integers
($1$--$10$); review-comment free-form text was not extracted or
used. ICLR submission titles and abstracts are author-curated and
pre-screened by the venue, so we did not apply additional
offensive-content filtering.
The empirical findings on review miscalibration are descriptive and
intended to inform program-committee design; they should not be used
to discount the work of individual reviewers, area chairs, or program
chairs, whose decisions are made under realistic load constraints that
we do not simulate. We caution against using a single disruption
metric (EDM, LLM rater, or otherwise) as a direct input to acceptance
decisions: per-paper ranks are pipeline-sensitive (see
Section~\ref{sec:discussion}), and the same structural bias toward
established research vocabularies that the metrics themselves exhibit
could be amplified by such use. The LLM rater is prompted with title
and abstract only and is not used to generate per-paper publishable
judgments.

\section*{Acknowledgements}

We thank the maintainers of the open-source datasets, models, and
software libraries on which this study depends, including the
\texttt{berenslab/iclr-dataset} of ICLR submissions and OpenReview
metadata \citep{berenslab2025iclr}, the Semantic Scholar Graph API for
citation data, the OpenAI text-embedding and gpt-4o-mini model
families, the Anthropic Claude family, the open-weight Llama-3.3
(Meta) and Qwen-2.5 (Alibaba) families, and the gensim,
scikit-learn, NetworkX, UMAP, HDBSCAN, and matplotlib libraries.
AI assistance was used only for grammar and style checks of the
manuscript text; all analyses, code, and figures were produced by the
authors.

\bibliography{custom}

\appendix

\section{Citation Network Quality Analysis}
\label{app:network}

\paragraph{ICLR submissions and acceptance.}
Table~\ref{tab:data} reports per-year ICLR submission counts and
acceptance outcomes. Accept rates are stable at $26\text{--}33\%$
across the decade, with $2017$ as an outlier at $40.5\%$ due to the
much smaller submission pool.

\begin{table}[ht]
  \centering
  \small
  \begin{tabular}{lrrr}
    \toprule
    Year & Submissions & Accepted & Accept \% \\
    \midrule
    2017 &    489 &    198 & 40.5 \\
    2018 &  1,012 &    336 & 33.2 \\
    2019 &  1,569 &    502 & 32.0 \\
    2020 &  2,593 &    687 & 26.5 \\
    2021 &  3,009 &    859 & 28.5 \\
    2022 &  3,422 &  1,094 & 32.0 \\
    2023 &  4,955 &  1,573 & 31.7 \\
    2024 &  7,401 &  2,261 & 30.5 \\
    2025 & 11,663 &  3,703 & 31.7 \\
    \midrule
    Total & 36,113 & 11,213 & 31.0 \\
    \bottomrule
  \end{tabular}
  \caption{ICLR submissions and acceptance outcomes per year.}
  \label{tab:data}
\end{table}

\paragraph{Semantic Scholar match rates.}
Table~\ref{tab:s2-match} reports the Semantic Scholar match rate by ICLR cohort year.
Matching uses title-based fuzzy similarity (RapidFuzz token-set ratio $\geq 0.85$)
with year tolerance $\pm 2$. Early cohorts (2017--2020) are almost fully matched
($>96\%$); match rates decline for 2021--2025 as newer papers have less complete
S2 indexing at retrieval time. Unmatched papers are concentrated among recent and
rejected submissions that were never posted to arXiv.

\begin{table}[ht]
\centering
\small
\begin{tabular}{lrrr}
\toprule
Year & ICLR papers & S2 matched & Match rate \\
\midrule
2017 &     489 &     483 & 98.8\% \\
2018 &   1,012 &     975 & 96.3\% \\
2019 &   1,569 &   1,553 & 99.0\% \\
2020 &   2,593 &   2,506 & 96.6\% \\
2021 &   3,009 &   2,396 & 79.6\% \\
2022 &   3,422 &   2,605 & 76.1\% \\
2023 &   4,955 &   3,692 & 74.5\% \\
2024 &   7,401 &   5,364 & 72.5\% \\
2025 &  11,663 &   8,269 & 70.9\% \\
\midrule
Total & 36,113 & 27,843 & 77.1\% \\
\bottomrule
\end{tabular}
\caption{Semantic Scholar match rate by ICLR cohort year (RapidFuzz token-set
  ratio $\geq 0.85$, year tolerance $\pm 2$). Early cohorts are matched
  at $96.3\%$--$99.0\%$; match rates drop for 2021--2025, driven by
  larger submission pools
  and more incomplete S2 preprint coverage at time of retrieval.}
\label{tab:s2-match}
\end{table}

\paragraph{PDF-extraction fallback (ICLR 2017 pilot).}
A PDF pipeline (\texttt{pymupdf4llm} + regex) was piloted on all 489 ICLR 2017
papers. PDF download success was 99\%; however, extracted reference strings were
unresolved (e.g., ``Vaswani et al., 2017'') and would require an additional
fuzzy-match step to map to S2 records, defeating the purpose of a fallback.
Given the 98.8\% S2 match rate for 2017, the marginal benefit did not justify
extending the pipeline to the full corpus.

\section{Method Implementation Details}
\label{app:methods-extended}

\paragraph{(M2) Generic Graph Embedding: node2vec details.}
We train standard (undirected) node2vec \citep{grover2016node2vec} on the
undirected version of $G$ with walk length $T = 160$, $R = 80$ walks per
node, embedding dimension $d = 100$, and skip-gram context window
$c = 5$. For each paper $i$ we compute
\begin{equation}
  \text{M2}_i = 1 - \frac{\bar{\mathbf{r}}_i \cdot \bar{\mathbf{c}}_i}
                         {|\bar{\mathbf{r}}_i|\,|\bar{\mathbf{c}}_i|}
  \label{eq:n2v}
\end{equation}
where $\bar{\mathbf{r}}_i$ is the mean embedding of $i$'s references and
$\bar{\mathbf{c}}_i$ is the mean embedding of $i$'s citers. The comparison
between M2 and M3 isolates the contribution of direction-aware walk
training: both pipelines share all hyperparameters and differ only in
whether the random walks respect citation direction.

\paragraph{(M3) EDM training objective.}
Following \citet{kim2026edm}, the random-walk training objective with
single-side (left-context-only) skip-gram is
\begin{align}
  \mathcal{J} &\approx \sum_{u \in V} \sum_{v \in A_c(u)}
               \kappa_u^{\text{in}} \log \Pr(v \mid u) \\
  \Pr(v \mid u) &= \frac{\exp(\fvec_v \cdot \pvec_u)}{\sum_{v' \neq u} \exp(\fvec_{v'} \cdot \pvec_u)}
\end{align}
where $A_c(u)$ is the set of antecedent papers within $c$ citation steps
and $\kappa_u^{\text{in}}$ is the in-degree weighting term that biases
walks toward well-cited antecedents. The EDM score is the cosine distance
between past and future vectors, $\EDM_i = 1 - (\fvec_i \cdot \pvec_i)/(|\fvec_i|\,|\pvec_i|)$
(Eq.~\ref{eq:edm} in the main text).

\paragraph{(M4) LLM rater details.}
We prompt \texttt{gpt-4o-mini} \citep{openai2024gpt4omini} to score each
paper on a 0--10 integer disruption-potential scale given title and
abstract. The system prompt defines disruption as ``ideas, methods, or
findings that cause other researchers to abandon prior directions and
build in a new direction'' and explicitly distinguishes it from quality,
novelty, and citation impact. Scoring is deterministic (temperature
$=0$) and runs in parallel with 25 workers. The full prompt is in
Appendix~\ref{app:llm-prompt}. M4 captures a purely \emph{semantic}
notion of disruption grounded in paper content, independent of its
citation-graph position.

\paragraph{Topic modeling: details and topic-share formula.}
We embed \texttt{title [SEP] abstract} (capped at $6{,}000$ characters)
with OpenAI \texttt{text-embedding-3-large} \citep{openai2024textembed}
at $1{,}024$ dimensions for all $36{,}113$ papers. Following the
standard BERTopic-style recipe, we cluster in a \emph{reduced} space
rather than on the raw high-dimensional embeddings: UMAP
\citep{mcinnes2018umap} projects to $50$ dimensions
($n_{\text{neighbors}} = 15$, $\text{min\_dist} = 0$, cosine metric),
then HDBSCAN \citep{campello2013density}
($\text{min\_cluster\_size} = 50$, Euclidean metric, EOM cluster
selection) clusters that 50-D representation. A separate UMAP-2D
projection ($\text{min\_dist} = 0.1$) is used for visualization only.
Clustering on the 50-D UMAP rather than the $1{,}024$-D raw embeddings
is substantially faster and produces comparable or better cluster
quality because UMAP explicitly preserves both local and global
topology. For each topic cluster $T$ and year $y$, the topic share is
\begin{equation}
  \text{share}(T, y) = \frac{|\{i : \text{topic}(i) = T, \text{year}(i) = y\}|}
                            {|\{i : \text{year}(i) = y\}|}
\end{equation}
Topic-level citation flows are aggregated as cross-topic edge counts per
year pair $(T_{\text{src}}, T_{\text{tgt}})$.

\paragraph{Validation evaluation.}
For ERS and LAS we report ROC-AUC and odds ratios from Firth's
penalized logistic regression \citep{firth1993bias} with a $10$-percentile
increase in the measure as the predictor unit. Firth penalization is
essential here because both ERS and LAS have low positive rates that
bias standard MLE.

\section{EDM Parameter Sensitivity}
\label{app:params}

We test EDM's robustness to walk length $T$ and embedding dimension $d$ by
running the pipeline with $T \in \{80, 160, 240\}$ at $d=100$ and
$d \in \{64, 100, 128\}$ at $T=160$ (five configurations, 1D sweeps around the
canonical $T=160, d=100$). For compute tractability we use $R=20$ walks per
node in the sweep versus $R=80$ in the canonical run.

\begin{figure*}[ht]
  \centering
  \includegraphics[width=\linewidth]{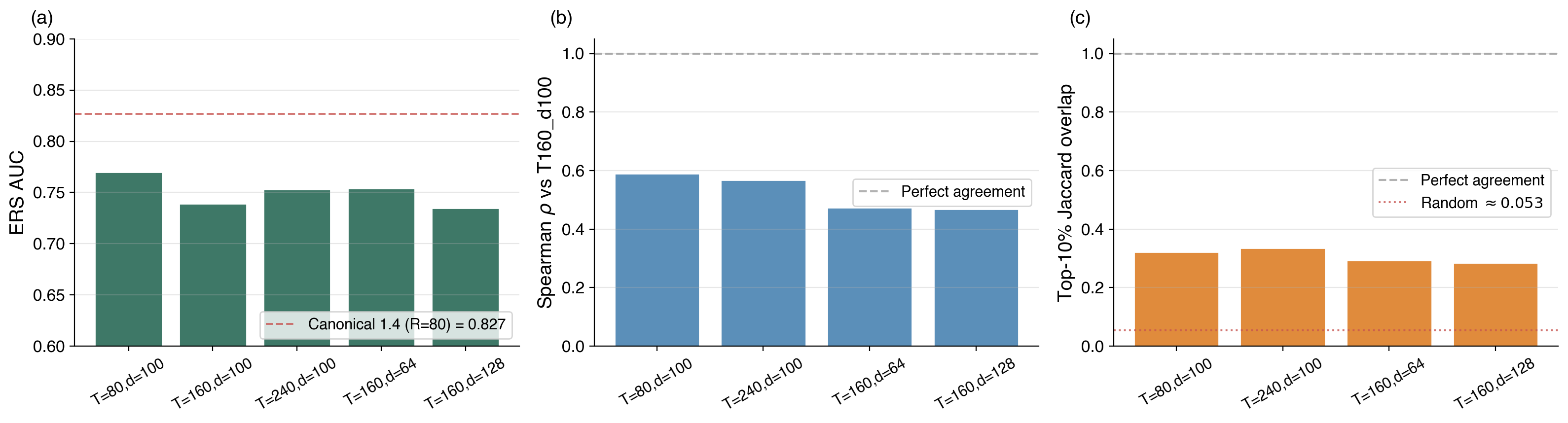}
  \caption{EDM parameter sensitivity. (a) ERS AUC across 5 configurations is
  tightly clustered in $[0.72, 0.76]$, within 6--9 points of the canonical
  $R=80$ run's 0.827 (red dashed line). (b, c) Pairwise Spearman $\rho$ and
  top-10\% Jaccard overlap of each sensitivity config against the $T=160,d=100,
  R=20$ sensitivity baseline are substantially above random.}
  \label{fig:sensitivity}
\end{figure*}

\paragraph{Validation quality is stable; exact ranks are not.}
All five sensitivity configurations achieve ERS AUC in $[0.72, 0.76]$, within
6--9 points of the canonical run's 0.827. The \emph{ability to identify
externally-recognized papers} is therefore robust to parameter choices.
However, exact paper rankings vary nontrivially: Spearman $\rho$ between
alternative configs and the sensitivity baseline ranges from 0.47 to 0.59.

\paragraph{Training pipeline affects ranks more than hyperparameters.}
A surprising finding is that the canonical ($R=80$, sequential walks) and
sensitivity ($R=20$, parallel walks) pipelines produce weakly correlated
individual rankings (Spearman $\rho \in [-0.18, -0.10]$ across the five
sensitivity configurations, mean $-0.145$), even at matched
hyperparameters. This
is because Word2Vec skip-gram training is sensitive to the order in which
sentences (walks) are presented: sequential generation produces ordered
batches while parallel generation via \texttt{imap\_unordered} randomizes
them, and this difference accumulates over 5 training epochs into
substantially different embeddings. We therefore treat EDM scores as
calibrated-within-pipeline but not guaranteed to reproduce exact ranks across
pipeline choices. \emph{Distribution-level conclusions} (top deciles, mean
trends, validation AUC) are robust; \emph{individual paper ranks} carry
pipeline noise on the order of $\rho \in [0.47, 0.59]$. All main-text results
use the canonical pipeline.

\paragraph{Multi-seed ensemble (rank-stability mitigation).}
We trained EDM under $K{=}3$ seeds (42, 7, 123) at the sensitivity-baseline
configuration ($R{=}20$, $T{=}160$, $d{=}100$, $c{=}5$, single-process
walk generation). Cross-seed pairwise Spearman $\rho = 0.55 \pm 0.05$
(range $[0.52, 0.61]$; $18{,}475$ papers with all-seed scores), and
top-$10\%$ Jaccard overlap $= 0.27$ across seed pairs. Per-seed ERS
AUC is $0.59\text{--}0.61$ (lower than the canonical $0.83$, as expected
from $R{=}20$ vs.\ $R{=}80$); the median-rank ensemble achieves AUC
$0.61$, modestly better than any single seed. The headline implication
is that the \emph{cross-pipeline} near-zero-$\rho$ result above is
specifically a pipeline-noise effect, not a seed-noise effect:
same-pipeline cross-seed stability is moderate, and a small-$K$
ensemble further attenuates per-paper rank variance.

\section{LLM Rater (M4): Prompt and Robustness Considerations}
\label{app:llm-prompt}

\paragraph{Model.} \texttt{gpt-4o-mini}, accessed via the OpenAI Chat
Completions API. Deterministic decoding ($\tau=0$), parallelism: 25
workers. Total inference: $36{,}113$ short calls.

\paragraph{System prompt (verbatim).}
\begin{quote}\small\ttfamily
You are a scientific reviewer evaluating the \emph{disruption potential} of
a research paper based only on its title and abstract.

Disruption is defined as: ``ideas, methods, or findings that cause
other researchers to abandon prior directions and build in a new
direction.'' A paper that consolidates an existing line of work (e.g.,
an improved benchmark, a careful empirical study) is \emph{not} disruptive.

Disruption is distinct from:
\begin{itemize}\item Quality. A paper can be excellent but consolidating.
\item Novelty. A novel contribution may still extend an existing line.
\item Citation count. A widely-cited paper may be consolidating; a
       lightly-cited paper may still be disruptive.\end{itemize}

Return a single integer from 0 (clearly consolidating) to 10 (paradigm-shifting),
with no additional text.
\end{quote}

\paragraph{User prompt.}
\texttt{Title: \{title\}\char`\\nAbstract: \{abstract\}}

\paragraph{Robustness considerations.}
The score depends on choice of LLM and on phrasing of the rubric. We
measure LLM-of-choice sensitivity directly via the cross-vendor
agreement study below; we did not run a multi-prompt ablation. Our
findings about EDM--LLM complementarity rely only on the LLM score
being \emph{any reasonable} semantic signal, not on its absolute
calibration; this is the safer interpretation supported by the data.

\paragraph{Inter-LLM agreement (vendor robustness).}
We re-scored a $2{,}000$-paper stratified sample
(year-group $\times$ decision $\times$ EDM tercile) with four
additional LLMs using the identical rubric: \texttt{gpt-4o} (OpenAI),
\texttt{claude-sonnet-4-6} (Anthropic),
\texttt{llama-3.3-70b-instruct} (Meta, via OpenRouter), and
\texttt{qwen-2.5-72b-instruct} (Alibaba, via OpenRouter). All five
models returned valid scores at $99.0\%$--$100\%$ rates. Across the
resulting $10$ pairwise comparisons, mean Spearman
$\rho = 0.60$ (range $[0.46, 0.72]$) and mean quadratic Cohen's
$\kappa = 0.28$ (Table~\ref{tab:llm-agreement}). This supports the
M4 claim: the LLM signal is rubric-bound, not gpt-4o-mini-specific,
and stable across four vendors (OpenAI, Anthropic, Meta, Alibaba).

\begin{figure*}[ht]
  \centering
  \includegraphics[width=0.85\linewidth]{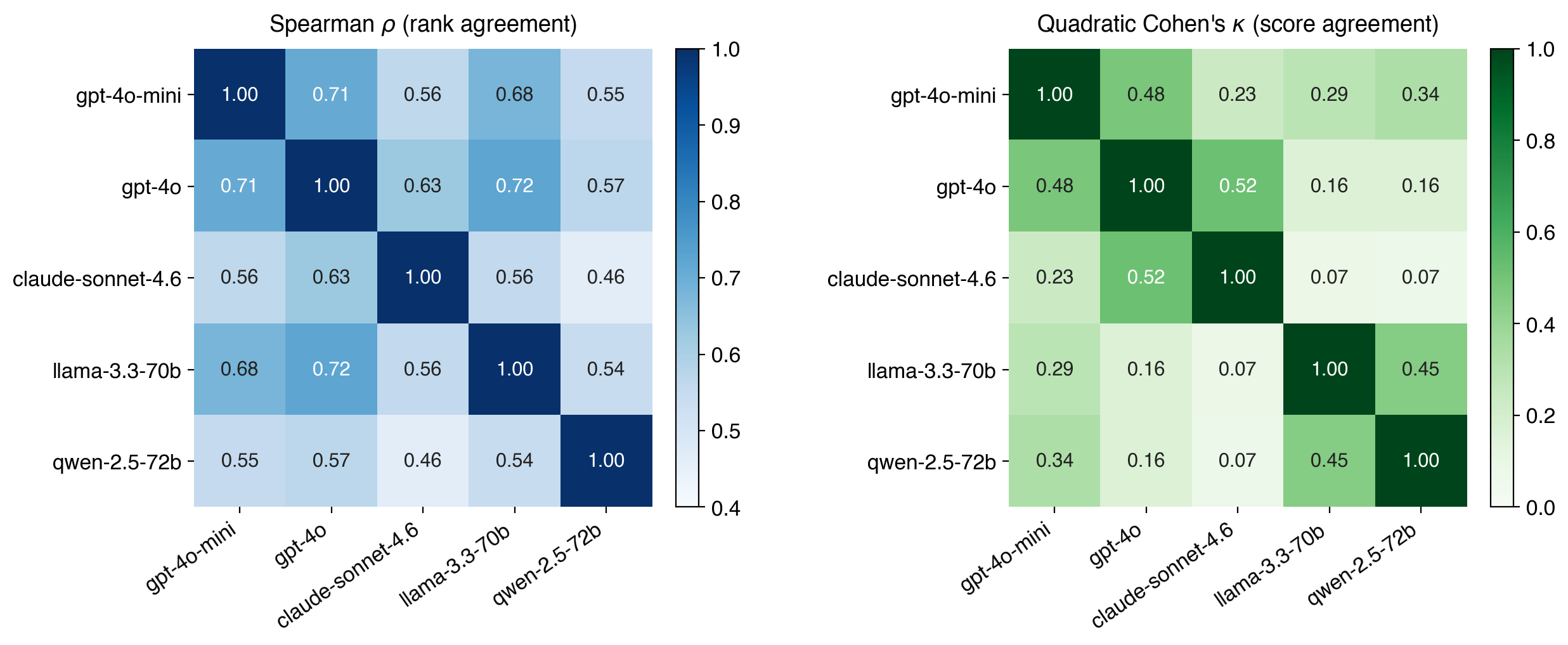}
  \caption{Inter-LLM agreement (heatmap counterpart to
  Table~\ref{tab:llm-agreement}). Left: Spearman $\rho$
  (rank agreement). Right: quadratic Cohen's $\kappa$ (score
  agreement). Off-diagonal mean $\rho = 0.60$ and $\kappa = 0.28$.
  The same-vendor pair (gpt-4o-mini versus gpt-4o, top-left) shows
  the strongest agreement, and cross-vendor pairs are uniformly
  above chance, which is consistent with the LLM rater signal
  being rubric-bound rather than model-specific.}
  \label{fig:llm-agreement-heatmap}
\end{figure*}

\paragraph{Note on Claude snapshot.}
An earlier attempt used the retired \texttt{claude-3-5-sonnet-20241022}
snapshot which returns $404$ Not~Found on current Anthropic accounts;
all calls failed silently in the wrapper and yielded $0$ valid scores
across $2{,}000$ papers. A small diagnostic pass with single-call probes
isolated the cause and confirmed that \texttt{claude-sonnet-4-6} returns
clean rubric-format responses without prefilling or special parsing.
All results in this section use that snapshot.

\paragraph{Scoring distribution.}
The full-corpus distribution of M4 scores is bimodal at $5$ and
$7$, with $7/10$ corresponding to the $83.8$th percentile and $8/10$ to
the $99.6$th. Among the $27$ ICLR Best Paper winners (2021--2025),
GPT-4o-mini assigns $7/10$ to most papers; only ``Score-Based
Generative Modeling through SDEs'' (2021) receives $8/10$. This
saturation is discussed in Appendix~\ref{app:rq1-extended}.

\section{Validation Set Details}
\label{app:validation}

\subsection*{External Recognition Set (ERS)}
The ERS consists of all ICLR papers in the top $2\%$ by ICLR-internal citation
count (citations received from other papers in the 36,113-paper corpus).
The 98th-percentile threshold is $33$ inbound citations; $739$ papers meet this
criterion ($739 / 36{,}113 = 2.05\%$, slightly above $2\%$ due to ties at the
boundary). The threshold adapts to the empirical distribution rather than a
fixed count; the ICLR-internal citation network is sparse ($12{,}831$ papers
have at least one citer), and a fixed count would produce too few positives
for reliable AUC estimation.

Papers without any ICLR-internal citations receive a count of zero and are
not ERS positives. The citation count baseline is excluded from ERS evaluation
because ERS is defined by citation count (making any AUC for that baseline
trivially circular).

\subsection*{LLM-judge Annotation Set (LAS)}
The LAS is a stratified random sample of $50$ papers labeled by two
independent runs of an LLM judge (\texttt{claude-opus-4-6}, Anthropic)
over each paper's title and abstract. Stratification was by:
\textit{(i)} year-group (early: 2017--2018; mid: 2019--2021; late:
2022--2024); \textit{(ii)} acceptance status (26 accepted, 24
rejected); and \textit{(iii)} CD-score tercile (low/mid/high) to avoid
over-sampling high-CD papers. This design ensures the sample is
representative of the corpus across time, status, and CD distribution.

Each run independently emitted a binary disruption label (1 =
disruptive, 0 = consolidating) and a confidence rating (1--3). Run~1
labeled 7 papers disruptive (14\%) and Run~2 labeled 4 (8\%). Under
\textit{union aggregation} (either run says disruptive), the LAS
contains $\mathbf{9}$ positive and $\mathbf{41}$ negative labels.
Under intersection aggregation (both agree), only $2$ papers are
positive, which is too few for reliable AUC estimation. Cohen's
$\kappa = 0.291$ (86\% raw agreement) measures run-to-run stochastic
consistency of the LLM judge, not human inter-rater reliability;
the standard Landis--Koch interpretive scale therefore does not
apply. The LAS provides a cross-model semantic-rubric baseline
(judge \texttt{claude-opus-4-6} vs.\ M4 rater \texttt{gpt-4o-mini}),
not a human-validated gold standard; we discuss the implications in
Section~\ref{sec:discussion}.

\subsection*{ICLR Best Paper Case Studies}
The best-paper set consists of 27 Outstanding Paper Award winners from ICLR
2021--2025 with per-year distribution 8/7/4/5/3 (2021/2022/2023/2024/2025),
matched by OpenReview forum ID from the official ICLR website and human-verified. These papers are used only for qualitative
interpretation and coverage comparisons (Appendix~\ref{app:rq1-extended}), not for
AUC or regression evaluation.

\section{LAS Annotation Protocol (LLM Judge)}
\label{app:annotation}

The LAS labels come from an LLM judge (\texttt{claude-opus-4-6},
Anthropic) prompted with the rubric below, run twice independently
over the same $50$-paper sample. The rubric was originally drafted to
instruct human annotators in an earlier design iteration; the wording
and field schema were preserved when the protocol moved to the LLM
judge.

\paragraph{Task.}
For each paper, the judge reads the title and abstract and assigns a
disruption label based on whether the paper primarily \emph{disrupts}
or \emph{consolidates} existing research directions:

\begin{itemize}
  \item Disruptive (1): The paper introduces ideas, methods, or findings
        that cause subsequent researchers to move away from prior work
        and build in a new direction. Citing papers tend to cite this
        paper \emph{instead of} its references.
  \item Consolidating (0): The paper extends, refines, or synthesizes
        existing work without fundamentally redirecting the field.
        Citing papers tend to cite this paper \emph{alongside} its
        references.
\end{itemize}

\paragraph{Key distinctions.}
The prompt explicitly states that (a) disruption is not the same as
quality, since a paper can be excellent but consolidating (for
example, a thorough benchmark); (b) disruption is not the same as
novelty, since a novel contribution may consolidate if it extends an
existing line; and (c) disruption is not the same as citation count,
since a highly cited paper may be consolidating and a lightly cited
paper may be disruptive.

\paragraph{Output fields.}
Each output row records three items per paper: a disruption label
($0$ or $1$), a confidence rating ($1=$ low, $2=$ medium, $3=$ high),
and optional free-text notes for borderline cases.

\paragraph{Provided context.}
Each prompt included the paper's CD disruption index, ICLR-internal
citation count, and acceptance status, marked as for reference only
with an explicit instruction not to copy the CD score. The judge had
access only to title, abstract, and these reference fields; it did
not retrieve the full paper.

\paragraph{Run-to-run aggregation.}
No adjudication round was conducted. The two runs were aggregated by
\emph{union} (positive if either run labeled the paper disruptive)
for LAS evaluation. Cohen's $\kappa = 0.291$ between the two runs is
a measure of run-to-run stochastic consistency of the judge under the
same rubric, not human inter-rater reliability; it primarily reflects
ambiguity in the rubric plus stochasticity in the judge's outputs.

\paragraph{Calibration examples in the prompt.}
The rubric provided four canonical calibration examples:
\emph{``Attention Is All You Need''} (disruptive, confidence~3,
introduced Transformers replacing RNN and CNN seq2seq models);
\emph{``BERT''} (disruptive, 3, shifted NLP toward pre-train and
fine-tune); \emph{``A Survey of Deep Learning for NMT''}
(consolidating, 3, synthesizes existing work); and \emph{``Improved
Regularization with Cutout''} (consolidating, 2, incremental
data-augmentation extension).

\section{RQ1 Extended Analyses}
\label{app:rq1-extended}

\begin{figure*}[ht]
  \centering
  \includegraphics[width=\linewidth]{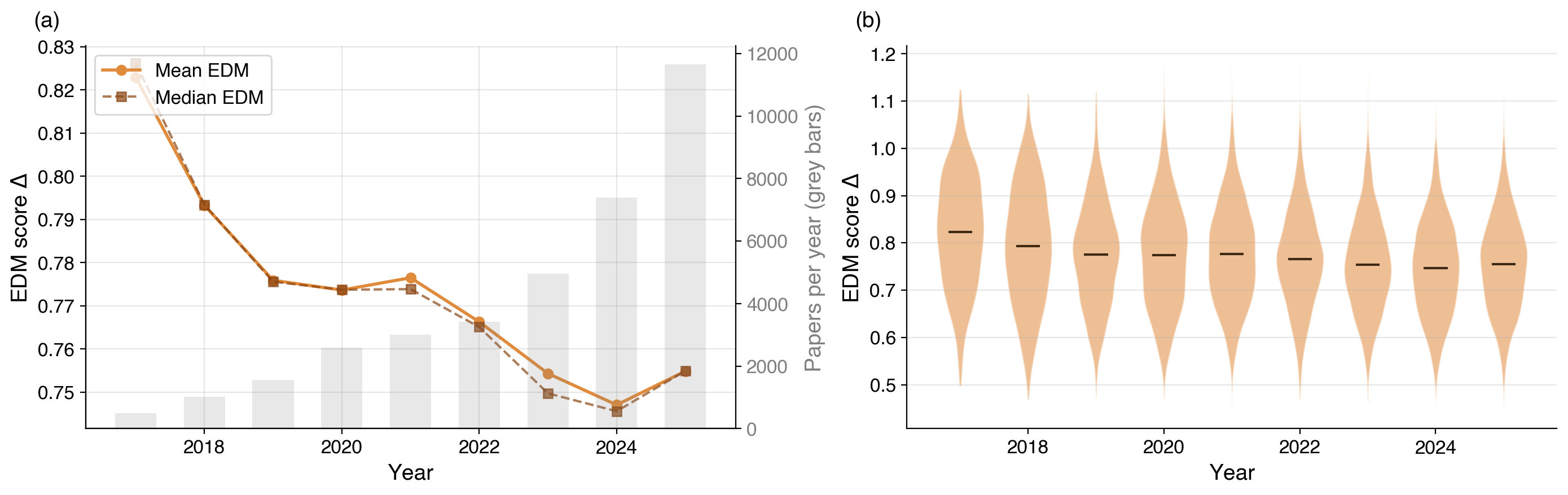}
  \caption{Temporal trend of EDM across ICLR 2017--2025.
  (a) Mean and median $\Delta$ are stable over the decade.
  (b) The year-wise violin plot shows that disruptiveness distributions
  shift modestly but do not exhibit the monotonic decline reported for
  Web-of-Science-indexed journal science \citep{park2023papers}.}
  \label{fig:yearly}
\end{figure*}

\paragraph{Temporal trend of EDM.}
Mean and median $\EDM$ at ICLR fluctuate modestly around $0.76$ across
$2017\text{--}2025$ without the monotonic decline reported by
\citet{park2023papers} for broad science (Fig.~\ref{fig:yearly}).
Possible interpretations: (i) the 9-year ICLR window is too short to
detect decadal change; (ii) AI's rapid iteration introduces both
consolidating benchmark work and disruptive paradigm shifts in
comparable proportions; (iii) ICLR's growth in submissions (489 in
2017 to $11{,}663$ in 2025) dilutes the signal from individual
disruptive papers.

\paragraph{Pairwise scatter and correlation.}

\begin{figure}[ht]
  \centering
  \includegraphics[width=\linewidth]{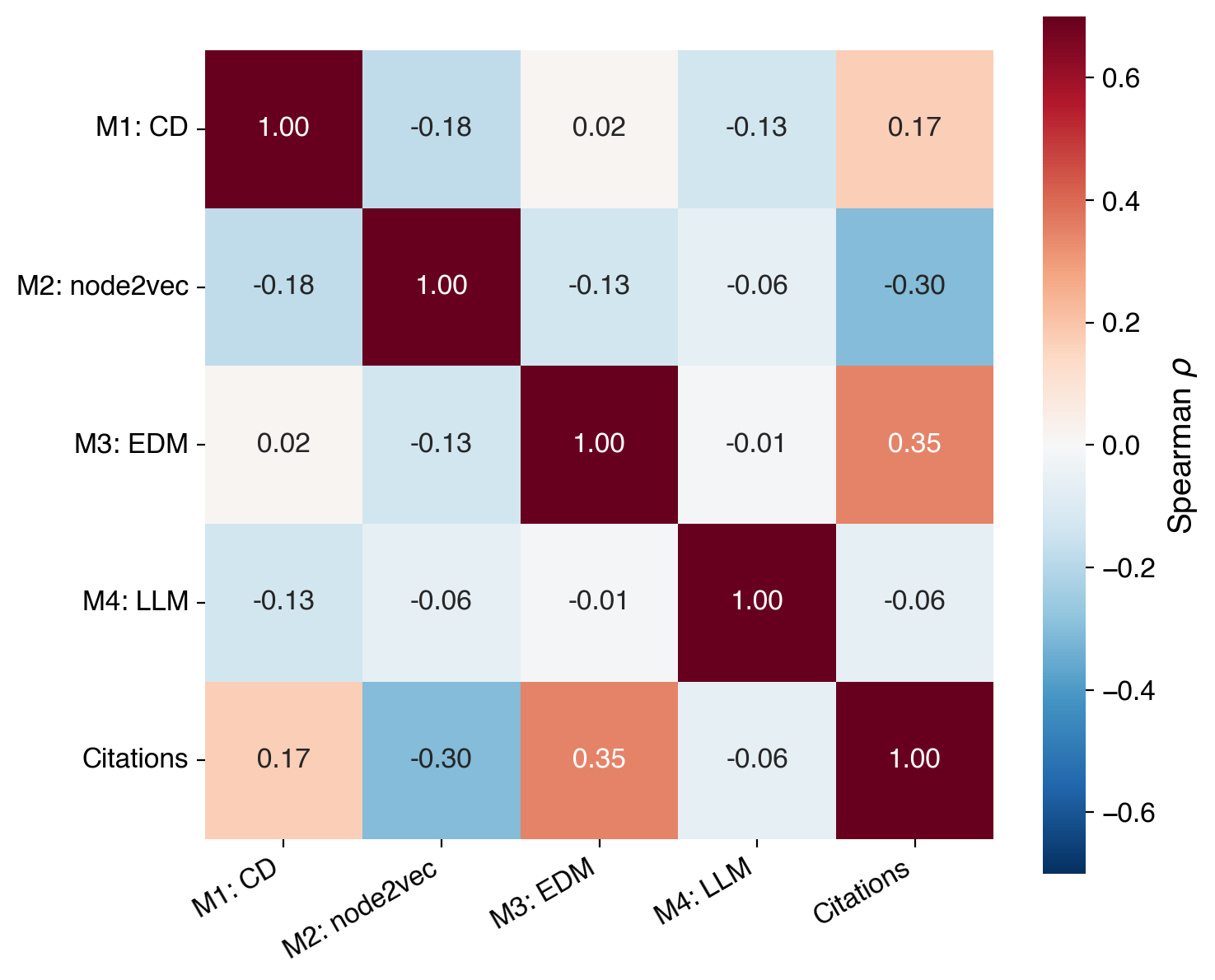}
  \caption{Spearman correlation between disruptiveness measures across
  ICLR 2017--2025. Near-zero correlations between M4 (LLM) and the
  citation-based measures (M1, M3) confirm that semantic and structural
  signals access distinct information channels.}
  \label{fig:corr}
\end{figure}

\begin{figure*}[ht]
  \centering
  \includegraphics[width=\linewidth]{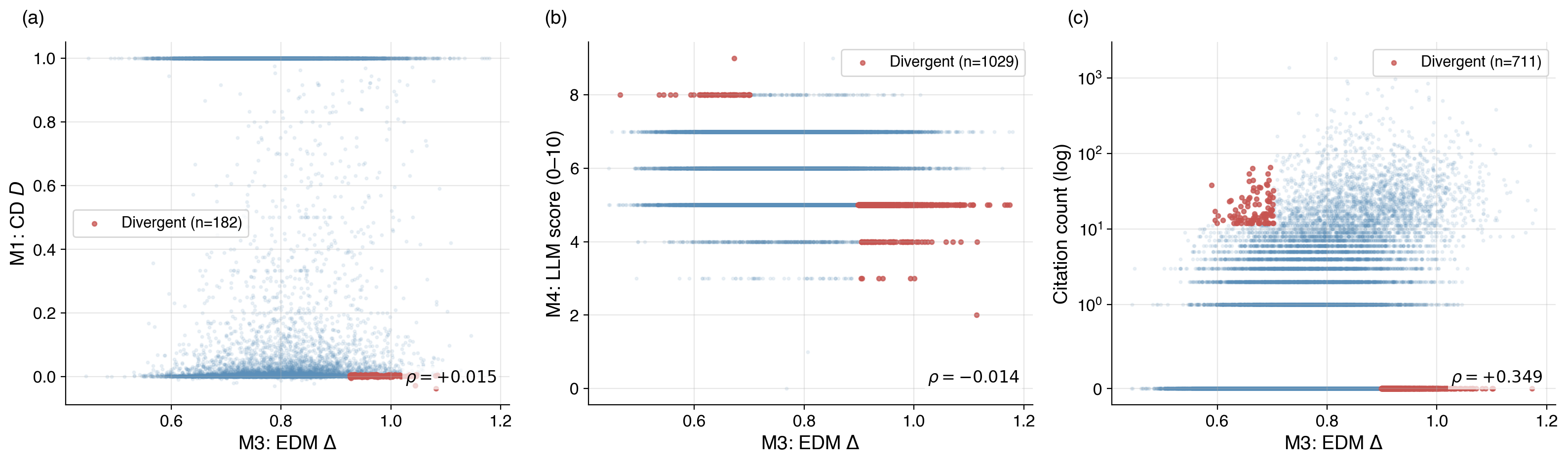}
  \caption{Pairwise scatter plots between M3 EDM and M1 CD (a), M4 LLM
  (b), and citation count (c). Red dots mark \emph{divergent} papers:
  top decile on one measure but bottom 30th percentile on the other.
  The divergent population between EDM and LLM is substantial,
  motivating pairing the two measures for downstream catalyst
  classification.}
  \label{fig:scatter}
\end{figure*}

On the $27$ ICLR Best Paper winners (2021--2025), we compute
per-measure coverage to test how each disruptiveness measure handles
recent, partly-cited work. M1 (CD) covers $14/27$, M2 (node2vec)
covers $12/27$, M3 (EDM) covers $24/27$, and M4 (LLM) covers $27/27$;
seven of the $8$ ICLR 2024--2025 winners fail both M1 and M2 because
the citation window is too narrow, while EDM produces a score
because its random walks pick these papers up through other
contexts. The coverage gap is a practical limitation of
citation-based disruption measures for recent work and a concrete
argument for pairing them with embedding-based or content-based
measures in fast-moving fields.

As a temporally-normalized check on the ERS ranking, we compute the
Spearman correlation between each measure and citations per year
since publication. EDM yields $\rho = +0.300$ ($p < 0.001$), CD
$\rho = +0.127$ ($p < 0.001$), node2vec $\rho = -0.293$
($p < 0.001$), and LLM $\rho = +0.016$ ($p = 0.074$, ns). EDM and
CD are correctly signed; node2vec is \emph{inversely} correlated
with impact velocity, confirming that it does not measure disruption
in any useful sense; LLM shows no significant association with
citation velocity, consistent with its independence from
citation-graph signals. The ranking EDM $\gg$ CD $\gg$ node2vec is
therefore consistent across both ERS (static) and citation velocity
(time-normalized), reducing the risk that the ERS result is an
artifact of how ERS is defined.

To identify EDM's failure modes on field-shaping work, we examine
the per-paper EDM percentiles of the Best Paper winners. Several
clearly field-shaping papers receive low EDM scores: Score-Based
Generative Modeling through SDEs (2021, EDM $26.4$th percentile),
Analytic-DPM (2022, $8.4$th), Generalization in Diffusion (2024,
$2.0$th), and Learning Mesh-Based Simulation (2021, $17.1$th). These
papers extended existing research lines (score-based SDEs built on
NCSN; Analytic-DPM refined DDPM) without introducing a structurally
new vocabulary, so when future citers remain close to the paper's
own antecedents the past and future vectors stay similar by
construction. We accordingly recommend pairing EDM with a semantic
measure (M4) that can flag conceptually important follow-ups graph
structure alone would miss.

To probe the LLM rater's ceiling, we examine the scores assigned by
GPT-4o-mini to the $27$ ICLR Best Paper Award winners (2021--2025).
The model assigns $7/10$ to most papers (corresponding to the
$83.8$th full-corpus percentile); only \emph{Score-Based Generative
Modeling through SDEs} (2021) receives $8/10$ ($99.6$th percentile),
and no winner receives $9/10$ or $10/10$. The LLM rater therefore
reliably distinguishes disruptive papers from consolidating ones but
saturates within the top tier and cannot differentiate among the
most exceptional contributions; this ceiling effect is complementary
to EDM's blind spot, since EDM \emph{under}-ranks canonical
follow-ups whereas the LLM \emph{cannot rank} among papers it
correctly identifies as outstanding.

\section{RQ2 Supplementary Analyses}
\label{app:rq2-extended}

\paragraph{Topic clustering parameters.} We use OpenAI
\texttt{text-embedding-3-large} with \texttt{dimensions}$=1024$. For
HDBSCAN we cluster on a 50-dim UMAP projection
($n_{\text{neighbors}}=15$, $\text{min\_dist}=0$, cosine) and require
$\text{min\_cluster\_size}=50$. This produces $113$ topics and leaves
$12{,}102$ papers ($33.5\%$) as noise. The ten largest topics cover
over $50\%$ of clustered papers (Table~\ref{tab:topics}); the remaining
$103$ topics cover niche areas (quantum ML, symbolic regression, causal
effect estimation, etc.).

\begin{table}[ht]
  \centering
  \small
  \begin{tabular}{rrp{4.9cm}}
    \toprule
    Topic & Size & Label (top keywords) \\
    \midrule
    33 & 1{,}844 & Neural Network Optimization \\
    72 & 1{,}072 & Diffusion-based Image Generation \\
    17 & 1{,}000 & Adversarial Robustness Techniques \\
    2  &   853 & Molecular Design and Generation \\
    81 &   822 & Multimodal Vision-Language Integration \\
    6  &   798 & Federated Learning Optimization \\
    62 &   704 & Graph Neural Networks \\
    14 &   611 & Physics-Informed Neural Networks \\
    3  &   573 & Continual Learning Strategies \\
    16 &   572 & Explainable AI Techniques \\
    \bottomrule
  \end{tabular}
  \caption{Ten largest topic clusters in ICLR 2017--2025 (of 113 total).
  Labels generated by GPT-4o-mini from each cluster's top TF-IDF keywords
  and representative papers.}
  \label{tab:topics}
\end{table}

\begin{figure}[ht]
  \centering
  \includegraphics[width=\linewidth]{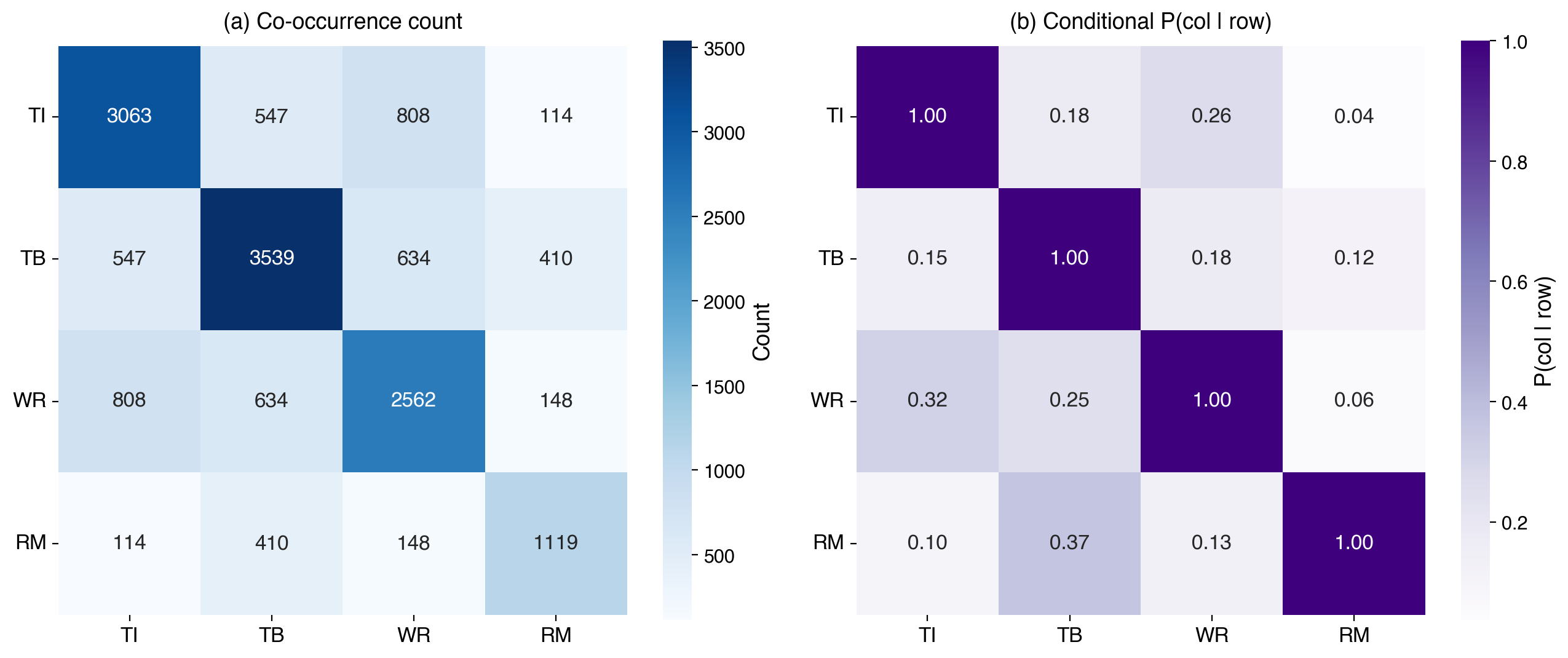}
  \caption{Catalyst co-occurrence. Left: raw counts of papers with both
  labels. Right: conditional $P(\text{column} \mid \text{row})$. The
  off-diagonal structure supports a multi-label rather than mutually
  exclusive catalyst formulation.}
  \label{fig:overlap}
\end{figure}

\paragraph{Catalyst type counts and co-occurrence.}
TB=$3{,}539$ ($9.8\%$), TI=$3{,}063$ ($8.5\%$), WR=$2{,}562$ ($7.1\%$),
RM=$1{,}119$ ($3.1\%$); union $=8{,}015$ ($22.2\%$).
Figure~\ref{fig:overlap} shows the full co-occurrence matrix.

\paragraph{Operational thresholds.} Each catalyst label is produced by a
thresholded operational criterion applied to the EDM and topic artifacts
(Table~\ref{tab:catalyst-thresh}).

\begin{table}[ht]
  \centering
  \small
  \begin{tabular}{lp{5.6cm}}
    \toprule
    Type & Threshold \\
    \midrule
    TI & Topic-share growth $\geq 2.0\times$ baseline \\
    TB & Top $10\%$ cross-topic flow and $\geq 2$ descendant topics \\
    WR & Centroid-shift $z$-score above cluster-conditional median \\
    RM & EDM $\geq$ 90th pct and borderline/contested review \\
    \bottomrule
  \end{tabular}
  \caption{Operational thresholds for catalyst label assignment.}
  \label{tab:catalyst-thresh}
\end{table}

\paragraph{TI: year-matched control growth.}

\begin{table}[ht]
  \centering
  \small
  \setlength{\tabcolsep}{4pt}
  \begin{tabular}{@{}lrrr@{}}
    \toprule
    Group & $n$ & Mean growth & Median \\
    \midrule
    TI papers              & 3{,}063 & 5.91         & 3.23 \\
    Year-matched controls  & 9{,}336 & 0.78         & 0.66 \\
    Ratio of means         & ---     & $7.55\times$ & --- \\
    \bottomrule
  \end{tabular}
  \caption{TI topic-share growth factor versus year-matched,
  different-topic controls (1--3 years after / 2 years before). Welch's
  $t = 46.06$, $p < 10^{-300}$. Controls' mean growth $< 1$ reflects
  background dilution of fixed-topic share as ICLR's submission volume
  grows; TI papers reverse this dilution.}
  \label{tab:ti-growth}
\end{table}

\begin{figure}[ht]
  \centering
  \includegraphics[width=\linewidth]{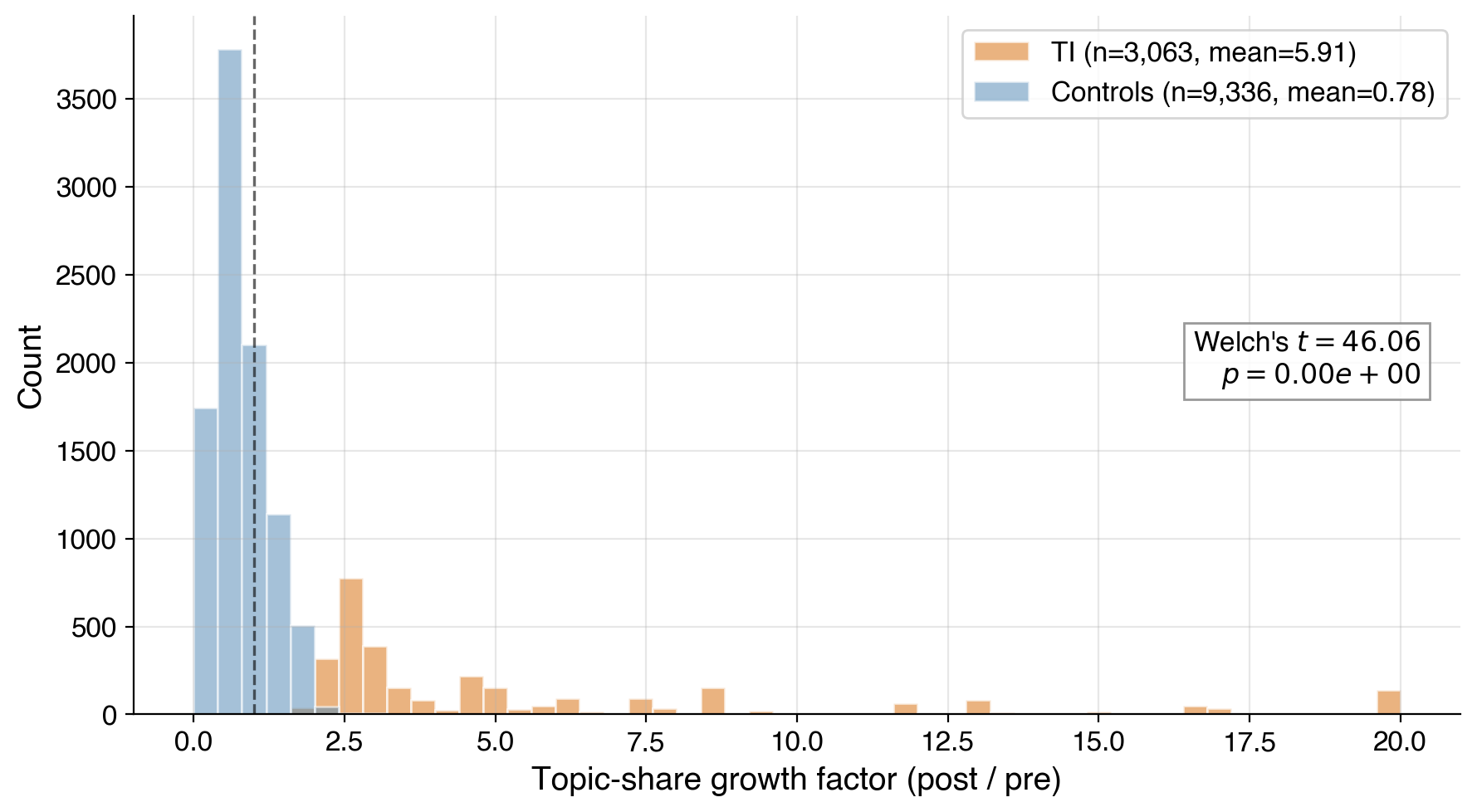}
  \caption{Distribution of topic-share growth factors for TI papers
  (orange) versus year-matched, different-topic controls (blue). The
  dashed line marks growth factor $= 1$ (no change). Controls cluster
  below 1 (dilution); TI papers cluster above 1 (growth).}
  \label{fig:ti-ctl}
\end{figure}

\paragraph{TB: cross-topic citation flow.}

\begin{table}[ht]
  \centering
  \small
  \begin{tabular}{lc}
    \toprule
    Metric & Value \\
    \midrule
    TB papers with valid pre+post windows & 3{,}228 \\
    Mean pre-window flow (edges/yr) & 242.3 \\
    Mean post-window flow (edges/yr) & 848.5 \\
    Mean growth factor & $11.52\times$ \\
    Median growth factor & $4.36\times$ \\
    25--75th percentile of growth & $[2.58, 9.34]$ \\
    \bottomrule
  \end{tabular}
  \caption{Cross-topic citation flow around Topic Bridge papers,
  measured as the ratio of mean inbound+outbound flow to the TB paper's
  topic cluster in the 1--3 years after publication versus the 2 years
  before.}
  \label{tab:tb-growth}
\end{table}

\begin{figure}[ht]
  \centering
  \includegraphics[width=\linewidth]{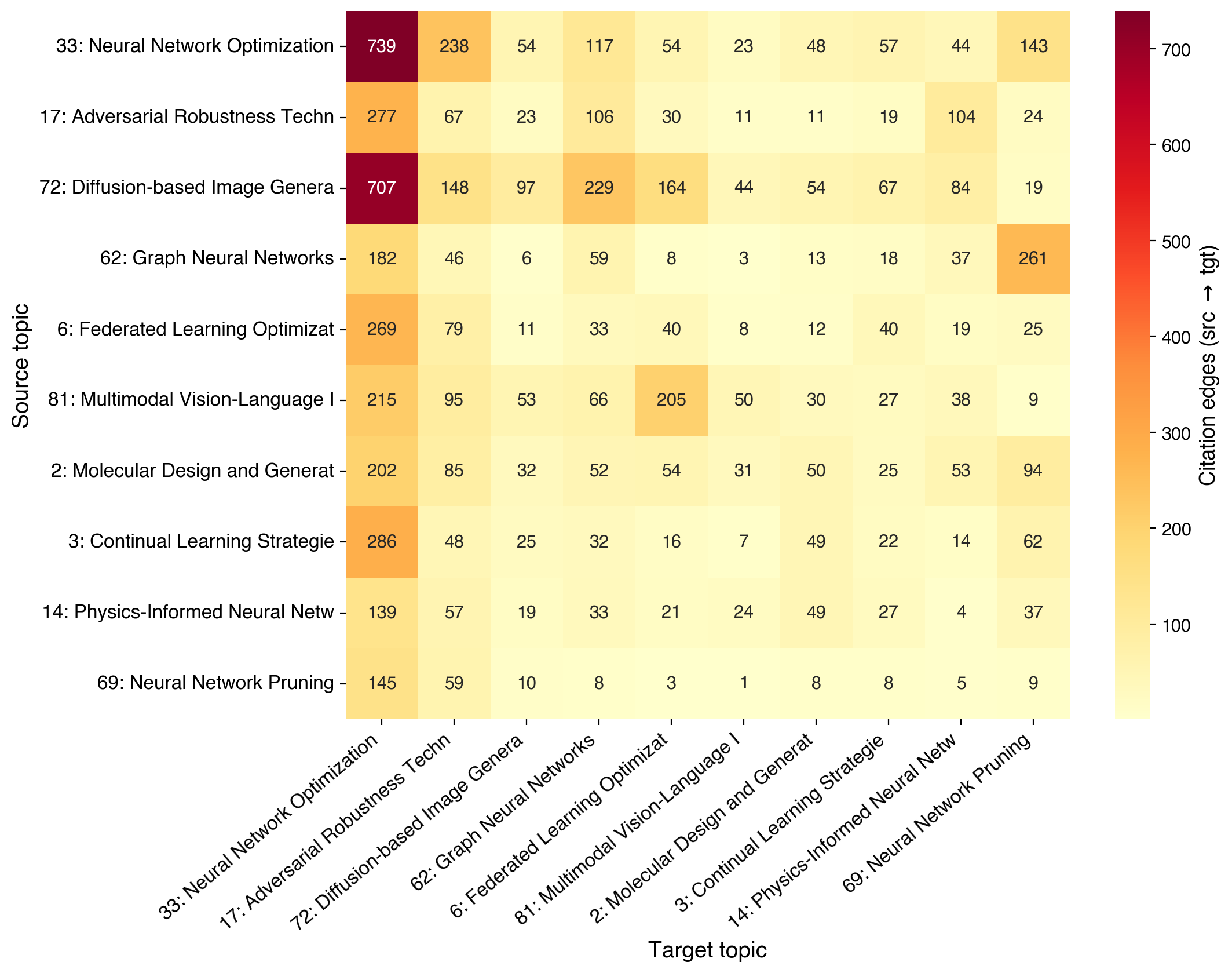}
  \caption{Cross-topic citation flow among the $10$ largest topic
  clusters. Cell $(i, j)$ counts source-target citation edges from
  topic $i$ to topic $j$. Adjacent topics (for example, diffusion to
  vision-language, or molecular to protein) show dense flow, while
  long-range pairs are sparser.}
  \label{fig:flow}
\end{figure}

\paragraph{Cross-topic flow records.} We aggregate ICLR-internal
citation edges by (source topic, target topic, year pair) into
$15{,}556$ flow records. TB papers are those whose publication is
followed within 2 years by an increase in inbound cross-topic flow to
or from their own topic cluster exceeding a per-pair baseline.

\paragraph{Simultaneous-discovery candidate funnel.}
Following \citet{kim2026edm}, we identify candidate pairs as papers
published in the same year whose future vectors have cosine
$\geq 0.9$ and no author overlap. The raw threshold produces
$483{,}809$ candidates from the $22{,}302$ papers with valid future
vectors. Inspection reveals many top-similarity pairs are artifacts of
sparse citation neighborhoods (the two highest-cosine pairs at $0.999$
similarity pair topically unrelated papers, e.g., ``Multi-Vector
Embedding on Networks with Taxonomies'' with ``Dynamic Least-Squares
Regression'', 2022). We apply three filters (Table~\ref{tab:sim-funnel}); the steep
drop from $482{,}850$ to $334$ at the citation-count filter shows that
the cosine threshold alone is dominated by under-determined pairs.

\begin{table}[ht]
  \centering
  \small
  \begin{tabular}{lr}
    \toprule
    Stage & Remaining pairs \\
    \midrule
    Raw candidates (cosine $\geq 0.9$) & 483{,}809 \\
    After no author overlap & 482{,}850 \\
    After both have $\geq 3$ internal citations & 334 \\
    After both in valid topic cluster & 162 \\
    \bottomrule
  \end{tabular}
  \caption{Filtering funnel for simultaneous-discovery candidates.}
  \label{tab:sim-funnel}
\end{table}

\paragraph{Citation-based validation of simultaneous discovery.}
We operationalize simultaneous discovery via two citation-graph
criteria: \emph{independence} (no edge $A \to B$ or $B \to A$) and
\emph{shared descendants} (co-citation rate
$|\mathrm{citers}(A) \cap \mathrm{citers}(B)| / \min(|\mathrm{citers}(A)|, |\mathrm{citers}(B)|) \geq 0.20$).

\begin{table}[ht]
  \centering
  \footnotesize
  \setlength{\tabcolsep}{3pt}
  \begin{tabular}{@{}lrrc@{}}
    \toprule
    Scope & Pass & Rate & 95\% CI \\
    \midrule
    Top-80 (cosine-ranked) & $61/80$   & $76.25\%$ & [65.4, 85.1]\% \\
    Filtered top-162       & $115/162$ & $71.0\%$  & [63.4, 77.8]\% \\
    Full 483{,}809 pool    & $975$     & $0.20\%$  & [0.19, 0.21]\% \\
    Direct-edge rate       & $562$     & $0.12\%$  & --- \\
    \bottomrule
  \end{tabular}
  \caption{Citation-based ``precision'' (sense of \citealp{kim2026edm}):
  no mutual citation and substantial co-citation overlap among
  descendants. Wilson 95\% CIs.}
  \label{tab:sim-precision-citation}
\end{table}

\paragraph{Threshold sensitivity.}
At the top-80, citation-based precision is $91.25\%$ under independence
alone ($73/80$), $85.0\%$ at co-citation rate $\geq 0.10$, $76.25\%$
at $\geq 0.20$, $62.5\%$ at $\geq 0.30$, and $46.25\%$ at $\geq 0.50$.
Precision is robust to threshold choice across this range. We report
$\geq 0.20$ as the primary number (at least one-fifth co-cited) without
forcing a hard constraint the pool cannot support.

\paragraph{Within-top ranking is not well-calibrated.}
Within the top-80, Spearman $\rho$(EDM cosine, co-citation rate)
$= +0.119$ ($p = 0.29$). The filter ``cosine $\geq 0.9$'' is effective
but finer cosine ranking within the top region does not predict
descendant structure. Precision is driven by the threshold, not the
ordering.

\paragraph{Confirmed same-topic pair.}
The single same-topic pair in the top-$80$, a 2023 robotics-manipulation
pair (\emph{``Toward Learning Geometric Eigen-Lengths Crucial for
Robotic Fitting Tasks''} and \emph{``A Massively Parallel Benchmark for
Safe Dexterous Manipulation''}), passes all thresholds up to
$\geq 0.30$ and is the first ICLR simultaneous-discovery pair confirmed
by our protocol.

\paragraph{Interpretation.}
Three non-exclusive explanations for why AI's simultaneous-discovery
rate is lower than the physics setting of \citet{kim2026edm}:
(i) arXiv preprint culture circulates ideas well before ICLR
deadlines, so parallel discovery in a journal-based field becomes
sequential citation in AI; (ii) ICLR's 9-year window contains 3--5
generations of research topics, so any simultaneous idea has time to
resolve into a citation hierarchy before both papers reach a
conference; (iii) papers often look similar because they target the
same benchmark, which reflects convergence on a predecessor rather
than independent discovery.

\paragraph{Cross-domain diffusion methodology.}
We retrieved citing works via the Semantic Scholar Graph API for all
$6{,}797$ catalyst papers with S2 identifiers plus a stratified
$2{,}000$-paper non-catalyst control (seed 42). For each cited paper
we kept only citing works within three years of its ICLR appearance
(up to $1{,}000$ per paper) and classified each as AI-core or non-AI
via S2's \texttt{fieldsOfStudy} tags. Of the $8{,}797$ target papers,
$7{,}332$ ($83.3\%$) yielded $\geq 1$ saved citing work and $1{,}463$
($16.6\%$) returned zero in-window citations on a confirmed retry at
$1$ req/s single-flight; these are treated as real zeros. The
collection yielded $792{,}018$ (cited, citing) pairs across
$140{,}073$ unique non-ICLR citers; $20.6\%$ of all citing works fall
outside AI-core.

\begin{figure}[ht]
  \centering
  \includegraphics[width=\linewidth]{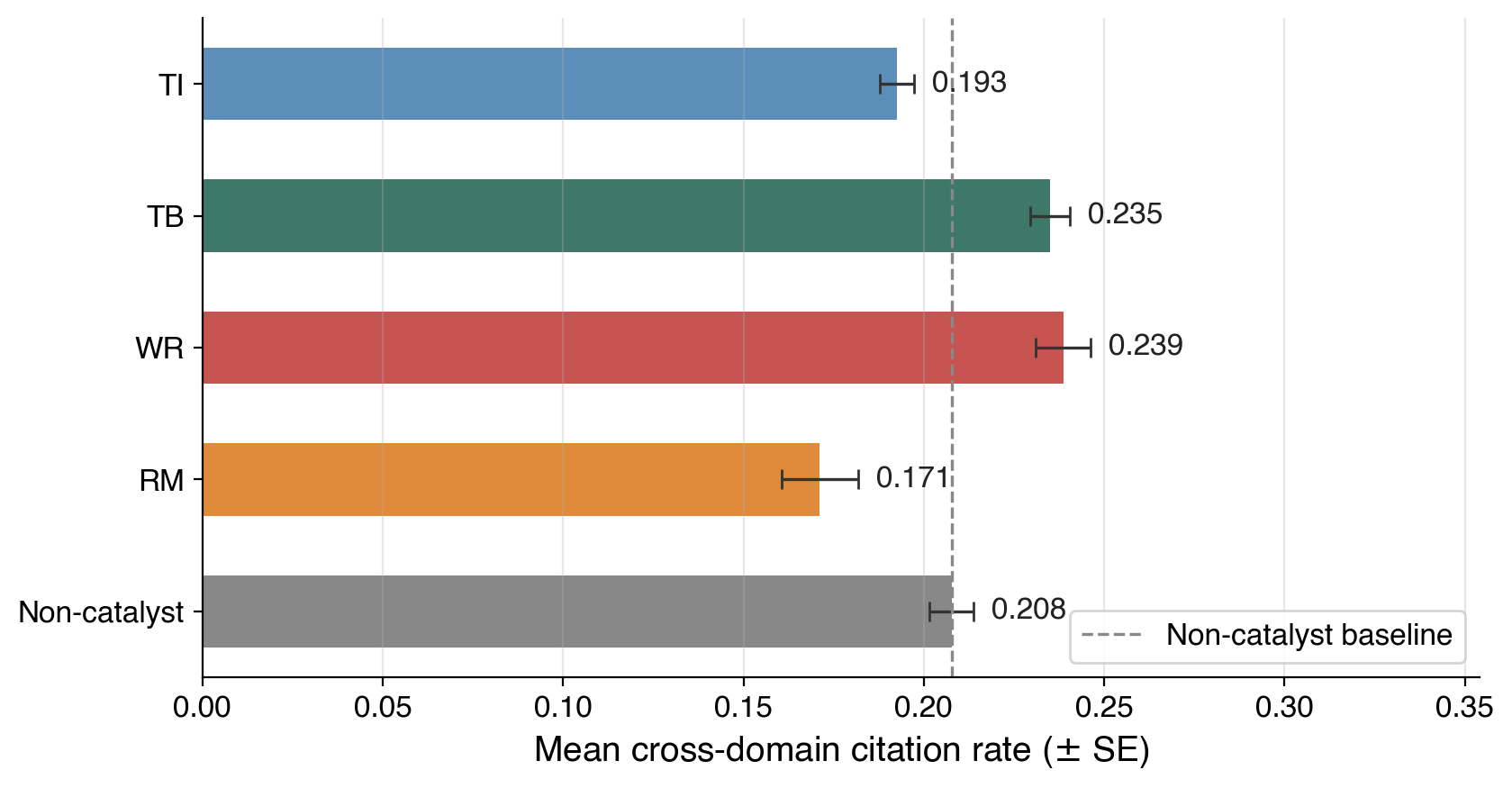}
  \caption{Mean cross-domain citation rate ($\pm$ SE) within three
  years of publication, by ICLR catalyst type.
  Rate~$=$~(non-AI)~/~(total). Dashed line: non-catalyst baseline.}
  \label{fig:cross-domain-rate}
\end{figure}

\begin{figure}[ht]
  \centering
  \includegraphics[width=\linewidth]{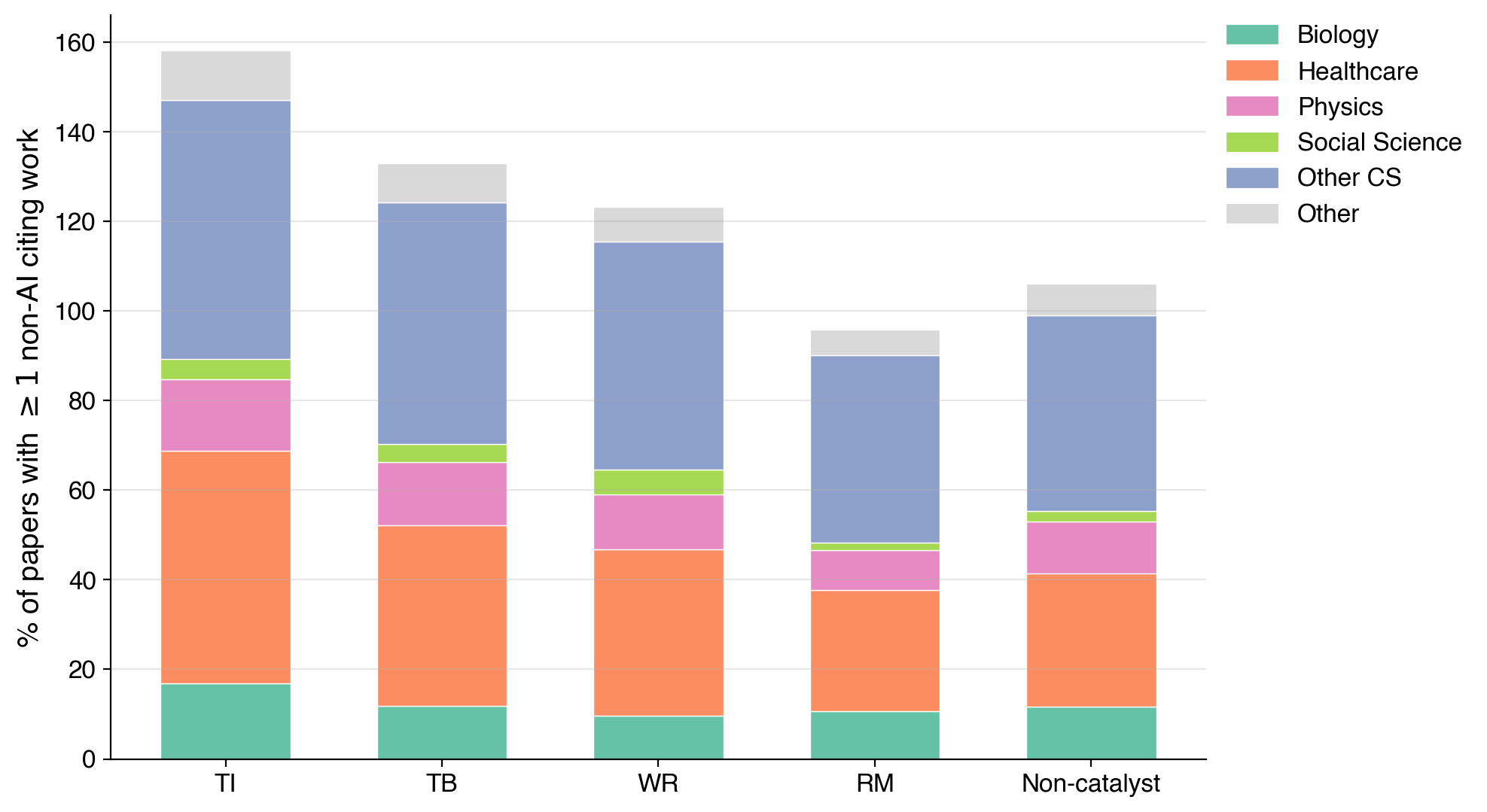}
  \caption{Fraction of papers with $\geq 1$ citing work in each non-AI
  domain, by catalyst type. Robotics is folded into Other CS (no
  standalone tag in S2).}
  \label{fig:cross-domain-domains}
\end{figure}

\paragraph{Composition vs.\ reach.}
Two complementary cross-domain metrics tell contrasting stories
(Figures~\ref{fig:cross-domain-rate} and~\ref{fig:cross-domain-domains}).
The mean \emph{composition} (fraction of each paper's citing set that
is non-AI) ranks TB and WR above the non-catalyst baseline, with TI
and RM \emph{below} (two-sided Mann--Whitney $p \leq 0.004$ for all
four types). The \emph{reach} metric (fraction of papers with
$\geq 1$ non-AI citing work) puts TI at the top of every non-AI
domain: $52\%$ Healthcare, $17\%$ Biology, $58\%$ Other-CS, exceeding
every other group including non-catalysts. TI papers accumulate large
total citation volumes that are AI-dominated in fraction but large
enough in absolute numbers to reach many external fields, while TB
and WR papers accumulate smaller footprints in which non-AI citations
make up a larger share.

\paragraph{Two-scale picture.} TB and WR are \emph{high-intensity,
narrow} catalysts: proportionally more cross-domain, consistent with
TB connecting topic clusters and WR generalizing methods adjacent
fields pick up. TI are \emph{low-intensity, wide}: absolute non-AI
footprint dominates every domain, but AI citation pull makes the
proportional signal small. RM are lowest on both metrics, reinforcing
the under-recognition story from RQ3.

\paragraph{Data limitations.} (i) The $10$-page S2 pagination cap
truncates citing-work lists at $1{,}000$ per paper; for the $5.8\%$
of papers that hit the cap, the saved sample is biased toward later
years of the three-year window (S2 returns citations newest-first).
(ii) The non-AI label follows S2's coarse \texttt{fieldsOfStudy}
taxonomy: papers in interdisciplinary venues tagged primarily
Computer Science are classified as AI-core even when their topical
focus is not. Both limitations push against our findings (they
suppress, not inflate, the measured cross-domain signal), so
reported effects are conservative.

\paragraph{TB threshold sensitivity.}
We swept the TB flow-threshold percentile $\in \{5\%, 10\%, 15\%, 20\%\}$ and
descendant-topic floor $\in \{2, 3, 5\}$ ($12$ configurations).
Mean post/pre flow growth ratio ranges $14.5\times$--$17.2\times$;
all $12$ configurations exceed $5\times$ (Table~\ref{tab:tb-sweep}).
The TB effect-size claim is therefore robust to threshold choice within a
$2\times$--$4\times$ window centered on the main-text canonical
($10\%$, $\geq 2$).

\begin{table}[ht]
\centering\footnotesize
\begin{tabular}{rrrr}
\toprule
flow \% & desc.\ floor & mean growth & median growth \\
\midrule
5  & 2 & 16.6 & 5.7 \\
5  & 3 & 16.7 & 5.7 \\
5  & 5 & 17.1 & 5.7 \\
10 & 2 & 16.3 & 5.7 \\
10 & 3 & 16.5 & 5.7 \\
10 & 5 & 17.2 & 5.7 \\
15 & 2 & 15.5 & 5.4 \\
15 & 3 & 15.9 & 5.6 \\
15 & 5 & 16.8 & 5.7 \\
20 & 2 & 14.5 & 5.4 \\
20 & 3 & 15.0 & 5.4 \\
20 & 5 & 16.3 & 5.7 \\
\bottomrule
\end{tabular}
\caption{TB threshold sweep ($12$ configurations). All exceed $5\times$
mean growth (range $14.5\times$--$17.2\times$). The main-text canonical
$(10\%, \geq 2)$ row reports $16.3\times$ on this full TB-flagged set;
the main-text headline figure $11.52\times$ is computed on the slightly
different ``valid pre+post window'' subset of TB papers
(Table~\ref{tab:tb-growth}).
}
\label{tab:tb-sweep}
\end{table}

\paragraph{Propensity-matched TI controls.}
To address the concern that TI papers might simply be more cited, we
re-ran the topic-share-growth analysis using $1$-NN propensity-matched
controls on $\log(1{+}\text{citation count})$, acceptance, and year
(caliper $0.05$ in propensity space; logistic-regression propensity
model with balanced class weighting). All $3{,}063$ TI papers matched
within caliper. The matched-control mean topic-share growth ratio is
$6.31\times$ (TI mean $5.88$, control mean $0.93$; Welch $t = 44.0$,
$p < 10^{-300}$; Mann--Whitney $p < 10^{-300}$). The $16\%$ drop from
the year-matched-only $7.55\times$ to the propensity-matched $6.31\times$
quantifies how much of the original effect is explained by citation-count
and acceptance differences; the residual effect remains large and highly
significant.

\paragraph{TI threshold sensitivity.}
A separate TI threshold sweep over growth ratios $\in [1.25, 3.0]$
under a simplified within-year control definition yields TI/control
ratios of $2.92\times$ to $3.24\times$, lower than the main-text
$7.55\times$. This reflects a methodological difference: the main-text
TI analysis uses year-matched, \emph{different-topic} controls
constructed in the main analysis pipeline, whereas the simplified
sweep uses year-matched, \emph{any-topic-with-non-TI-growth} controls
(a wider pool that includes within-topic non-TI papers whose topic
shares were already growing for unrelated reasons). The qualitative
finding that TI papers precede topic-share growth at a multiplicative
rate above $1$ holds in both formulations; the $7.55\times$ figure
depends specifically on the different-topic control design.

\section{RQ3 Supplementary Analyses}
\label{app:rq3-extended}

\begin{figure}[ht]
  \centering
  \includegraphics[width=\linewidth]{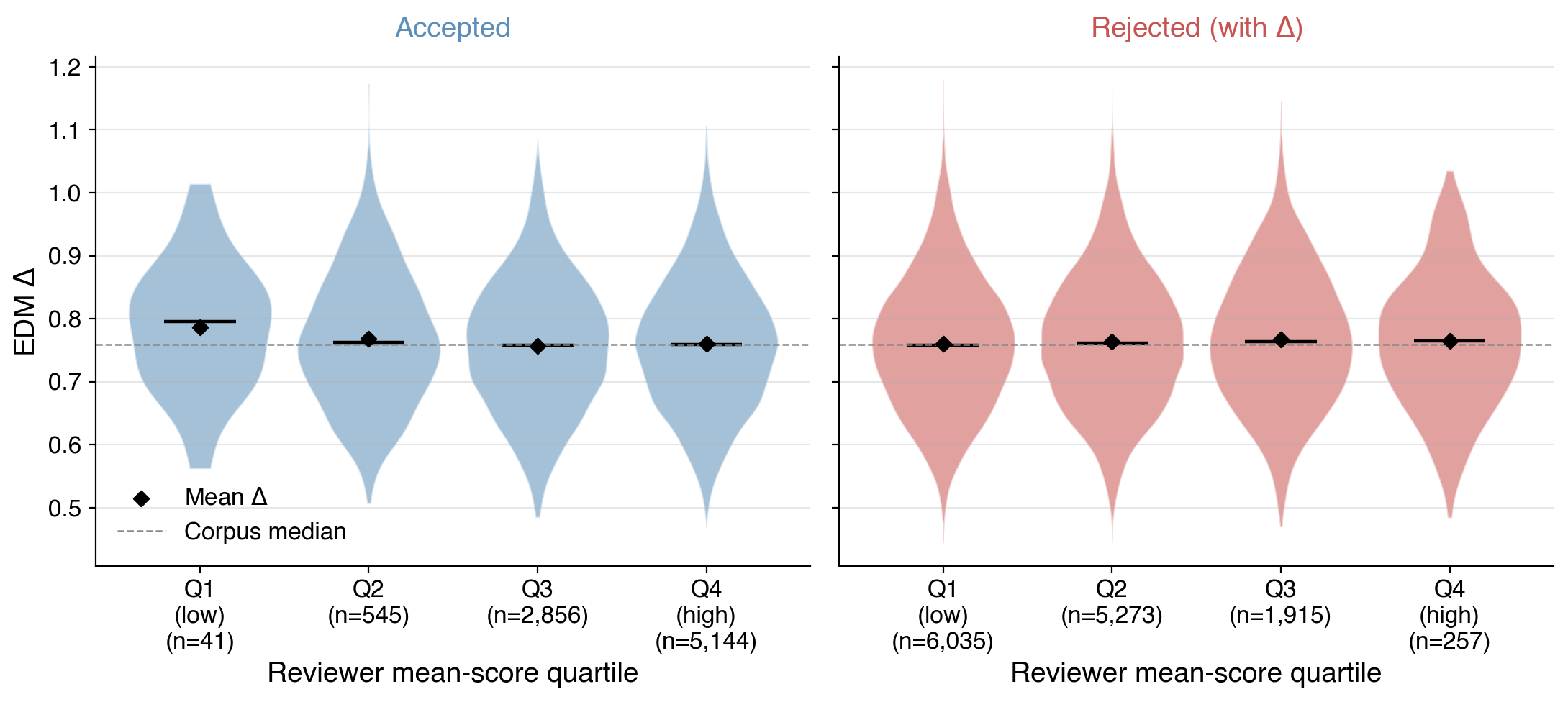}
  \caption{Distribution of EDM~$\EDM$ by reviewer mean-score quartile
  (accepted vs.\ rejected). Distributions are nearly identical across
  quartiles, consistent with the null correlation results in main-text
  Table~\ref{tab:firth-review}.}
  \label{fig:rq3-violin}
\end{figure}

\begin{figure}[ht]
  \centering
  \includegraphics[width=\linewidth]{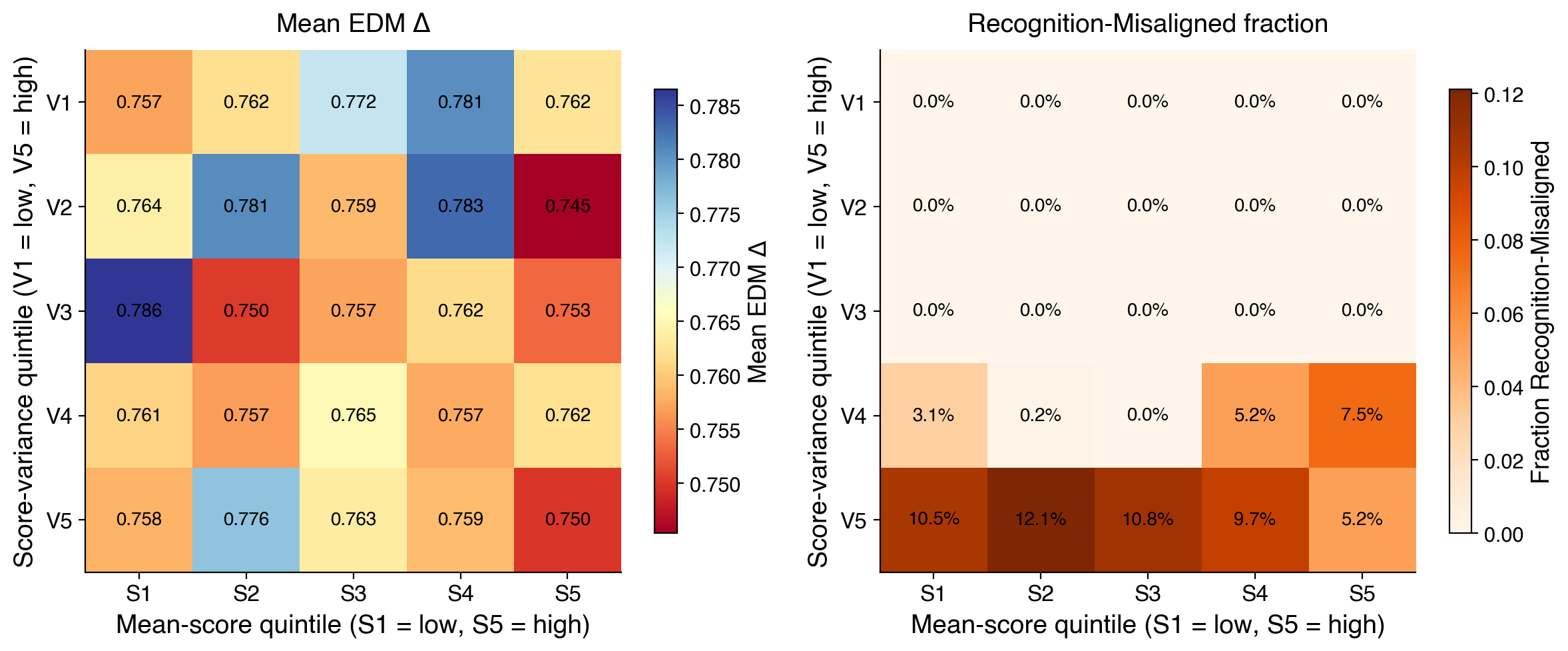}
  \caption{Mean EDM~$\EDM$ by reviewer mean-score bin and score-variance
  bin. Recognition-Misaligned and Topic-Bridge papers cluster in cells
  where review scores underestimate future disruption.}
  \label{fig:rq3-heatmap}
\end{figure}

\begin{table}[ht]
\centering
\footnotesize
\resizebox{\linewidth}{!}{%
\begin{tabular}{lrrrr}
\toprule
Type & $n$ & Gap & $t$ & $p$ \\
\midrule
Non-catalyst          & 15{,}723 & $-0.103$ & ---   & ---      \\
TI (Topic Initiator)  &  2{,}129 & $-0.095$ &  0.89 & $0.37$   \\
TB (Topic Bridge)     &  3{,}386 & $+0.145$ & 33.97 & $<0.001$ \\
WR (Within-topic Red.)&  1{,}708 & $-0.003$ &  9.16 & $<0.001$ \\
RM (Recog.-Misaligned)&  1{,}119 & $+0.590$ & 85.75 & $<0.001$ \\
\bottomrule
\end{tabular}}
\caption{Catalyst-type miscalibration. Mean review gap
($\EDM$ pctile $-$ score pctile) by catalyst type. Positive =
under-valued. $t$ vs.\ non-catalyst baseline.}
\label{tab:miscal-type}
\end{table}

\begin{table}[ht]
\centering
\footnotesize
\resizebox{\linewidth}{!}{%
\begin{tabular}{p{3.2cm}rr@{\hspace{1em}}p{3.2cm}rr}
\toprule
\multicolumn{3}{c}{Over-valued (trendy)} &
\multicolumn{3}{c}{Under-valued (niche)} \\
Topic & Gap & $n$ & Topic & Gap & $n$ \\
\midrule
Text-to-Video Gen.       & $-0.265$ & 123 & Quantum ML            & $+0.193$ & 35  \\
Masked Image Modeling    & $-0.260$ &  44 & Active Learning       & $+0.093$ & 66  \\
Vision Transformers      & $-0.249$ & 106 & Safe RL               & $+0.090$ & 59  \\
Optimal Transport        & $-0.241$ &  94 & Text Embeddings       & $+0.073$ & 86  \\
State Space Seq.\ Mod.   & $-0.225$ &  51 & Recommendation Sys.   & $+0.073$ & 64  \\
Diffusion Image Gen.     & $-0.220$ & 687 & Efficient Sampling    & $+0.070$ & 45  \\
3D Gen.\ w/Diffusion     & $-0.218$ &  79 & Meta-/Few-Shot Learn. & $+0.066$ & 209 \\
Energy-Based Gen.\ Mod.  & $-0.216$ &  44 & Continual Learning    & $+0.058$ & 348 \\
Gen.\ Flow Matching      & $-0.216$ & 107 & Cross-ling.\ Align.   & $+0.052$ &  68 \\
Backprop.\ Alternatives  & $-0.207$ &  50 & Federated Learning    & $+0.050$ & 477 \\
\bottomrule
\end{tabular}}
\caption{Topic bias. Top-10 over-valued and
under-valued topic clusters by mean review gap (EDM-$\EDM$
percentile minus reviewer-score percentile). Negative = over-valued
(reviewer scores exceeded future disruption); positive =
under-valued.}
\label{tab:topic-bias}
\end{table}

\begin{figure*}[ht]
  \centering
  \includegraphics[width=0.9\linewidth]{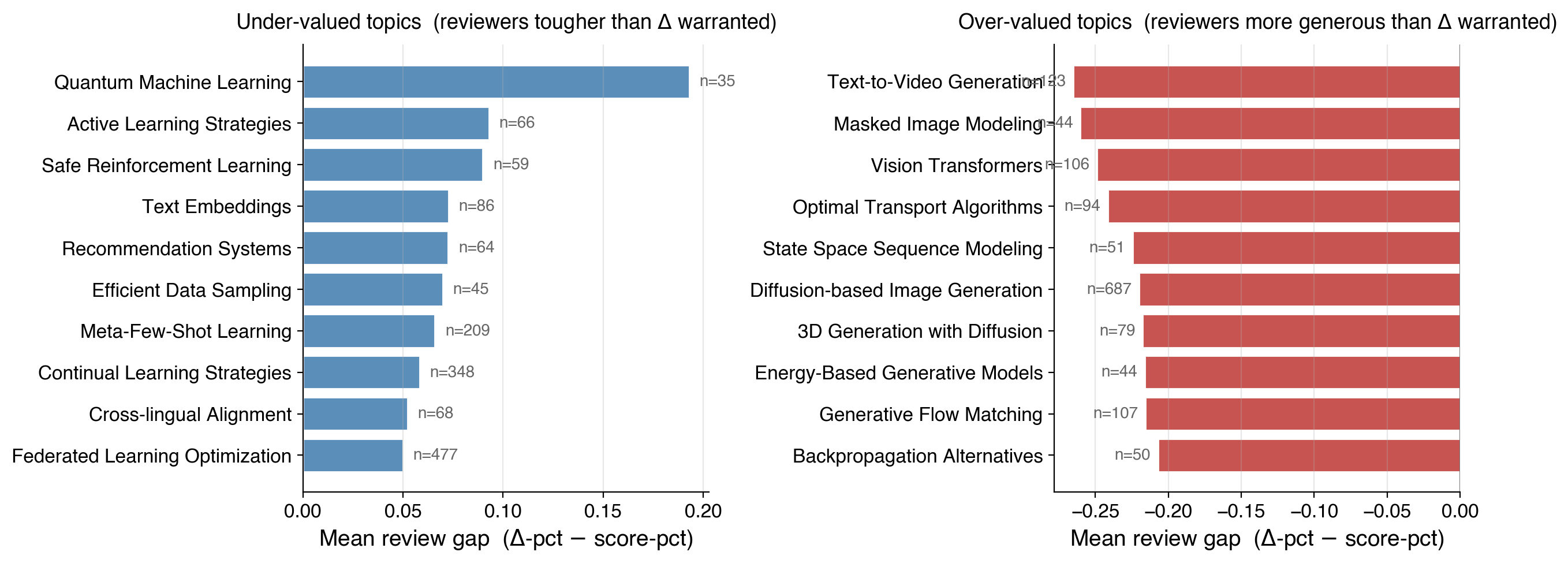}
  \caption{Mean review gap by topic cluster (top-10 over-valued and
  under-valued). Trendy sub-fields are systematically over-scored;
  niche and interdisciplinary areas are under-scored relative to their
  measured future-disruption contribution.}
  \label{fig:rq3-topic}
\end{figure*}

\begin{figure*}[ht]
  \centering
  \includegraphics[width=0.9\linewidth]{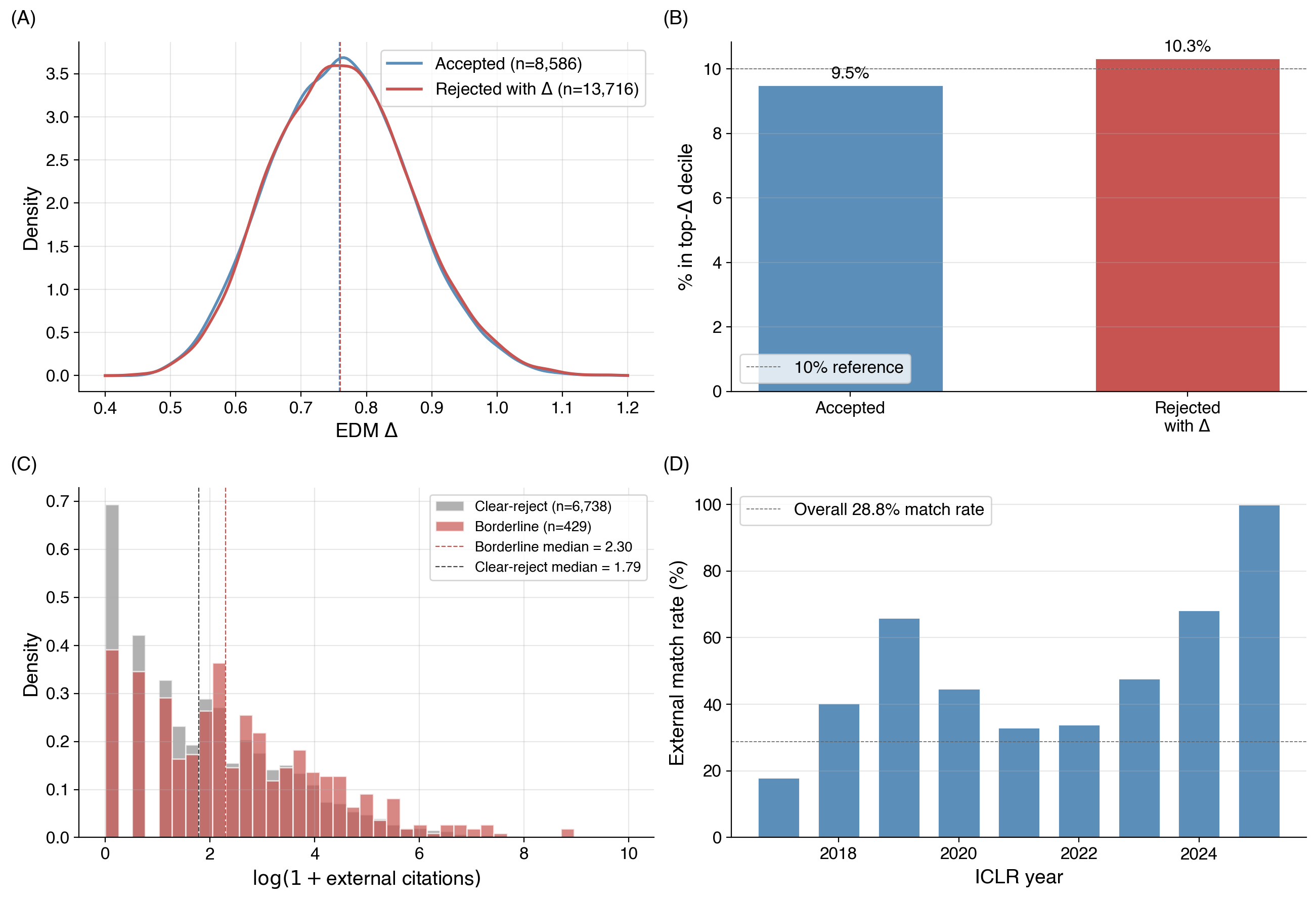}
  \caption{Rejected ICLR papers: trajectory and external-record analysis.
  (A) KDE of EDM ($\EDM$) for accepted versus rejected-with-$\EDM$
  papers. Distributions overlap closely (Mann--Whitney $p=0.11$).
  (B) Fraction of each group in the top-$\EDM$ decile
  ($\geq 0.899$); rejected papers are slightly over-represented
  ($10.3\%$ versus $9.5\%$).
  (C) External-citation distribution ($\log(1+\text{citations})$) for
  borderline-rejected versus clear-rejected papers among those matched
  in external records; borderline median $2.30$ versus clear-reject
  median $1.79$ (Mann--Whitney $p<0.001$).
  (D) External match rate by ICLR year; overall $28.8\%$ matched and
  $71.2\%$ unrecovered.}
  \label{fig:rq3-rejected}
\end{figure*}

\begin{table}[ht]
\centering
\footnotesize
\resizebox{\linewidth}{!}{%
\begin{tabular}{rrrrrrr}
\toprule
$\delta$ & $n_{\text{bord.}}$ & $n_{\text{clear}}$ & Bord.\ med. & Clear med. & Diff & $p$ \\
\midrule
$0.25$          &    $179$ & $6{,}988$ & $2.485$ & $1.792$ & $+0.693$ & $<0.001$ \\
$0.50$ &    $441$ & $6{,}726$ & $2.303$ & $1.792$ & $+0.511$ & $<0.001$ \\
$0.75$          &    $800$ & $6{,}367$ & $2.398$ & $1.792$ & $+0.606$ & $<0.001$ \\
$1.00$          & $1{,}457$ & $5{,}710$ & $2.398$ & $1.609$ & $+0.788$ & $<0.001$ \\
\bottomrule
\end{tabular}}
\caption{Borderline vs.\ clear-reject median $\log(1{+}\text{cit.})$ across threshold
  $\delta$ (score pts below year median-accepted; $n=7{,}167$ matched rejects).
  The $\delta=0.50$ row is the main-text definition. All $p<0.001$
  (one-sided Mann--Whitney).}
\label{tab:borderline-sensitivity}
\end{table}

\begin{table}[ht]
\centering
\footnotesize
\resizebox{\linewidth}{!}{%
\begin{tabular}{lrrrrc}
\toprule
Cutoff & $\EDM$ thr. & Acc.\ (\%) & Rej.\ (\%) & Diff & Dir. \\
\midrule
Top-10\% & $0.899$ & $9.5$ & $10.3$ & $+0.83$\,pp & $\checkmark$ \\
Top-5\%  & $0.941$ & $4.8$ & $5.2$  & $+0.41$\,pp & $\checkmark$ \\
Top-1\%  & $1.015$ & $0.8$ & $1.1$  & $+0.25$\,pp & $\checkmark$ \\
\bottomrule
\end{tabular}}
\caption{Top-$\EDM$ tail representation: accepted vs.\ rejected-with-$\EDM$
  ($n_{\text{acc}}=8{,}586$; $n_{\text{rej}}=13{,}716$). Over-representation
  is monotonically stable and never reverses.}
\label{tab:topdelta-robustness}
\end{table}

\section{Discussion: ICLR Setting and Program-Committee Implications}
\label{app:discussion-extended}

\paragraph{ICLR as a laboratory.}
Three features of ICLR are essential for science-of-science work and
unavailable in journal-based corpora. First, the OpenReview platform
provides numeric reviewer scores and acceptance decisions for every
submission, enabling direct measurement of reviewer calibration against
long-run trajectory change, an analysis that is not feasible in
journal-based science where review is private. Second, the dense
arXiv preprint culture gives rejected papers a mechanism to remain in
the scholarly record, producing the rare opportunity to measure
false-negative rates on a well-defined sample ($n=24{,}900$
rejections), of which $71\%$ disappear, providing a baseline for how
much potential disruption the ML and NLP community does not directly
observe. Third, the compressed timescales mean that within a nine-year
window, citation chains long enough to estimate EDM have already
formed for papers published in 2017, which allows disruption to be
studied near-prospectively rather than only retrospectively.

\paragraph{Peer review and recovery.}
The disruption-blindness of ICLR review persists even through repeated
review cycles: among the $280$ papers rejected at ICLR and later
accepted at a subsequent ICLR cycle, median $\EDM$ is $0.758$ vs.\
$0.759$ for never-rejected papers ($p{=}0.61$); the field's own
re-evaluation does not filter on disruptiveness either. The
rejected-paper analysis adds a sobering baseline: $71\%$ of rejected
papers disappear from the scholarly record entirely, so the
recoverable false-negative population is far smaller than the raw
rejection count implies. Borderline
rejections ($\log(1{+}\text{cit})$ median $2.30$ vs.\ $1.79$ for
clear rejects) represent a tractable intervention target for
program-committee design.

\paragraph{Actionable interventions for program committees.}
The structured nature of the miscalibration suggests concrete
interventions: (i) weight dissenting reviews more heavily for
submissions that bridge multiple topic clusters (the catalyst type with
the largest review-time miscalibration after RM); (ii) apply
topic-specific score adjustments to correct for documented
over-valuation of trendy areas (diffusion, ViT, SSMs) and
under-valuation of niche areas (quantum ML, federated learning,
safe RL); (iii) flag borderline-rejected papers for arXiv-cohort
follow-up to identify high-impact false negatives. All three
interventions are testable in OpenReview-instrumented future cycles.

\section{Per-Measure Limitations and Robustness Analysis}
\label{app:limitations-extended}

The body-text Discussion and Limitations section summarizes the main
limitations of the present analysis. This appendix records detailed
bounds on each measure and validation step.

\paragraph{Citation network coverage.} The citation graph is restricted
to ICLR-internal edges, constructed from Semantic Scholar reference
lists with a $77.1\%$ match rate against the ICLR corpus. Submissions
rejected at ICLR and never reposted to arXiv or another indexed venue
are not observed. Cross-disciplinary catalyst papers that enter ICLR
from outside the ICLR community are not represented by the
within-network measures (M1, M2, M3) and rely on M4 alone.

\paragraph{Coverage and recency.} M1 (CD) and M2 (node2vec) require
ICLR-internal citers and produce undefined scores for papers without
any in-corpus citers. This restricts their coverage to $35\%$ and
$18\%$ of the corpus respectively, and they fail on most 2024--2025
ICLR Best Paper Award winners. M3 (EDM) covers $62\%$ and M4 (LLM) covers
the full corpus. For early-impact detection, only M3 and M4 are usable
in practice.

\paragraph{EDM rank instability.} The validation AUC of M3 EDM is stable
across walk-length $T \in \{80, 160, 240\}$ and embedding dimension
$d \in \{64, 100, 128\}$ (ERS-AUC in $[0.72, 0.76]$;
Appendix~\ref{app:params}). Per-paper rank stability decomposes into
two regimes. Cross-pipeline (sequential vs.\ parallel walk generation
at matched hyperparameters) yields rankings statistically
indistinguishable from independent, because Word2Vec skip-gram
training is sensitive to the order in which walks are presented.
Cross-seed within-pipeline (three seeds at the App-C sensitivity
configuration) yields Spearman $\rho = 0.55 \pm 0.05$ with top-decile
Jaccard $0.27$; a median-rank three-seed ensemble preserves the
validation AUC. Distribution-level conclusions are therefore robust,
but per-paper EDM ranks should not be interpreted across pipelines
without an ensemble.

\paragraph{EDM canonical-follow-up blind spot.} By construction, EDM
penalizes papers whose intellectual descendants stay close to the
paper's own antecedents. This downweights canonical follow-ups,
including Score-Based Generative Modeling through SDEs and
Analytic-DPM, which sit in the bottom quartile of EDM despite being
widely recognized as field-shaping. The companion LLM rater M4 captures
this class through content rather than citation topology, and we
recommend reporting both measures jointly in any downstream use.

\paragraph{LAS statistical power.} The LLM-judge Annotation Set (LAS)
contains $50$ papers labeled by two independent runs of an LLM judge
(\texttt{claude-opus-4-6}; run-to-run $\kappa = 0.291$). Union
aggregation yields $9$ positive labels; intersection aggregation
yields $2$, which is insufficient for stable ROC-AUC estimation. The
M4 LLM odds ratio of $1.41$ ($p = 0.03$) is therefore reported with
wide confidence intervals. A larger validation pool with additional
independent LLM judges, and ideally a human-annotated gold standard,
is the most consequential extension for the semantic-rubric
agreement comparison.

\paragraph{LLM rater single-model dependence.} Full-corpus M4 scoring
uses a single model (\texttt{gpt-4o-mini}). Inter-LLM agreement on a
stratified $2{,}000$-paper subsample across five models from four
vendors (\texttt{gpt-4o-mini}, \texttt{gpt-4o}, \texttt{claude-sonnet-4-6},
\texttt{llama-3.3-70b}, \texttt{qwen-2.5-72b}) yields mean pairwise
Spearman $\rho = 0.60$ (Appendix~\ref{app:llm-prompt},
Table~\ref{tab:llm-agreement}). A multi-prompt ablation and a
systematic study of cross-vendor scoring bias are left to future
work.

\paragraph{Simultaneous-discovery validation.} The citation-graph
criterion for simultaneous discovery requires both candidate papers to
have a non-trivial descendant-citer base. Sparse-citer pairs cannot be
reliably evaluated by this protocol, so the reported precision applies
to the densely-cited subset of candidates rather than to the raw
cosine-ranked pool.

\end{document}